\documentclass[sigconf,balance=false]{acmart}
\usepackage{popets}
\usepackage{booktabs}
\usepackage{multirow}
\usepackage{tikz}
\usepackage{amsmath}
\usepackage[binary-units]{siunitx}
\usepackage[abbreviations]{foreign}
\usepackage{xspace}
\usepackage[inline]{enumitem}
\usepackage[algo2e,ruled,lined,boxed,commentsnumbered, noend, linesnumbered, procnumbered]{algorithm2e}
\usepackage{multicol}

\usepackage{pifont}%

\usepackage{ifthen}

\usepackage{acronym}
\acrodef{DL}{decentralized learning}
\acrodef{ML}{machine learning}
\acrodef{D-PSGD}{decentralized parallel stochastic gradient descent}
\acrodef{FL}{federated learning}
\acrodef{SGD}{stochastic gradient descent}
\acrodef{IID}{independent and identically distributed}
\acrodef{non-IID}{non independent and identically distributed}
\acrodef{RMSE}{root mean square error}
\acrodef{RMW}{random model walk}
\acrodef{GL}{gossip learning}
\acrodef{EL}{epidemic learning}
\acrodef{DWT}{discrete wavelet transform}
\acrodef{FFT}{fast Fourier transform}
\acrodef{MI}{mutual information}
\acrodef{DP}{differential privacy}
\acrodef{VN}{virtual node}
\acrodef{RN}{real node}
\acrodef{LDP}{local differential privacy}
\acrodef{PNDP}{pairwise network differential privacy}
\acrodef{PNLDP}{pairwise network local differential privacy}
\acrodef{GI}{gradient inversion}
\acrodef{CML}{collaborative machine learning}
\acrodef{TPR}{true positive rate}
\acrodef{FPR}{false positive rate}

\acrodef{LA}{linkability attack}
\acrodef{GIA}{gradient inversion attack}
\acrodef{MIA}{membership inference attack}
\acrodef{AIA}{attribute inference attack}
\acrodef{ROC}{receiver operating characteristic}
\acrodef{AUC}{area under the ROC curve} %
\newcommand{\sys}{\textsc{Shatter}\xspace}

\newcommand{\DL}{\ac{DL}\xspace}

\newcommand{\EL}{\ac{EL}\xspace}
\newcommand{\ROG}{\textsc{ROG}\xspace}

\newcommand{\topk}{\textsc{TopK}\xspace}

\newcommand{\cifar}{CIFAR-10\xspace}
\newcommand{\twitter}{Twitter Sent-140\xspace}
\newcommand{\movielens}{MovieLens\xspace}
\newcommand{\resnet}{ResNet-18\xspace}
\newcommand{\lenet}{LeNet\xspace}

\newcommand{\dpsgd}{{\xspace}\ac{D-PSGD}\xspace}

\newcommand{\niid}{\ac{non-IID}\xspace}

\newcommand{\muffliato}{{\xspace}\textsc{Muffliato}\xspace}

\graphicspath{ {figures/} }
\usepackage{tikz}
\usepackage[eulergreek]{sansmath}
\usepackage{pgfplots}
\usepackage{pgfplotstable}
\usepackage{cleveref}
\usepackage{comment}
\crefname{assumption}{assumption}{assumptions}

\pgfplotsset{compat=newest}
\usepgfplotslibrary{external,units,colorbrewer,groupplots,fillbetween,statistics}
\tikzsetexternalprefix{figures/}
\tikzset{external/mode=list and make}
\usetikzlibrary{patterns,shapes.misc}

\makeatletter
\begingroup\endlinechar=-1\relax
\everyeof{\noexpand}%
\edef\x{\endgroup\def\noexpand\homepath{%
		\@@input|"kpsewhich --var-value=HOME" }}\x
\makeatother

\newcommand{\inputplot}[2]{%
	\includegraphics{main-figure#2.pdf}
}

\newcommand{\newgroupwidth}[2]%
{\expandafter\xdef\csname groupwidth#1\endcsname{#2}}

\newcounter{groupwidth}
\newsavebox{\groupwidthbox}
\makeatletter
{\edef\groupnumber{#1}%
	\stepcounter{groupwidth}%
	\@ifundefined{groupwidth\thegroupwidth}{\pgfmathsetlengthmacro{\mywidth}{\linewidth/\groupnumber}}%
	{\expandafter\let\expandafter\mywidth\csname groupwidth\thegroupwidth\endcsname}%
	\begin{lrbox}{\groupwidthbox}%
		\tikzset{/pgfplots/width={\mywidth}}%
		\ignorespaces}%
	{\end{lrbox}%
	\usebox\groupwidthbox
	\pgfmathsetlengthmacro{\mywidth}{\mywidth + (\linewidth - \wd\groupwidthbox)/\groupnumber}
	\immediate\write\@auxout{\string\newgroupwidth{\thegroupwidth}{\mywidth}}}
\makeatother
\usepackage{amsmath,amssymb,amsfonts,amsthm}
\usepackage{physics}
\usepackage{enumitem}
\usepackage{thm-restate}
\usepackage{bbm}
\theoremstyle{definition}
\newtheorem{definition}{Definition}[section]
\theoremstyle{remark}
\newtheorem{remark}{Remark}
\newtheorem{theorem}{Theorem}
\newtheorem{lemma}[theorem]{Lemma}

\newtheorem{assumption}{Assumption}

\allowdisplaybreaks

\newcommand{\cC}{\mathcal{C}}
\newcommand{\cO}{\mathcal{O}}

\newcommand{\cE}{\mathcal{E}}
\newcommand{\cG}{\mathcal{G}}
\newcommand{\R}{\mathbb{R}}
\newcommand{\cV}{\mathcal{V}}

\newcommand{\esp}[1]{\mathbb{E}\left[#1\right]}
\newcommand{\proba}[1]{\mathbb{P}\left(#1\right)}

\setcopyright{popets}
\copyrightyear{YYYY}
\acmYear{YYYY}
\acmVolume{YYYY}
\acmNumber{X}
\acmDOI{XXXXXXX.XXXXXXX}
\acmISBN{}

\acmConference{Proceedings on Privacy Enhancing Technologies}
\begin{document}

\title{Noiseless Privacy-Preserving Decentralized Learning}

\author{Sayan Biswas}
\affiliation{%
  \institution{EPFL, Switzerland}
  \streetaddress{}
  \city{}
  \state{}
  \country{}
}

\author{Mathieu Even}
\affiliation{%
  \institution{Inria, DI ENS, PSL University, France}
  \streetaddress{}
  \city{}
  \state{}
  \country{}
}

\author{Anne-Marie Kermarrec}
\affiliation{%
  \institution{EPFL, Switzerland}
  \streetaddress{}
  \city{}
  \state{}
  \country{}
}

\author{Laurent Massouli\'e}
\affiliation{%
  \institution{Inria, DI ENS, PSL University, France}
  \streetaddress{}
  \city{}
  \state{}
  \country{}
}

\author{Rafael Pires}
\affiliation{%
  \institution{EPFL, Switzerland}
  \streetaddress{}
  \city{}
  \state{}
  \country{}
}

\author{Rishi Sharma}
\affiliation{%
  \institution{EPFL, Switzerland}
  \streetaddress{}
  \city{}
  \state{}
  \country{}
}
\authornote{Corresponding author: \texttt{first.last}@epfl.ch}

\author{Martijn de Vos}
\affiliation{%
  \institution{EPFL, Switzerland}
  \streetaddress{}
  \city{}
  \state{}
  \country{}
}

\renewcommand{\shortauthors}{Biswas et al.}

\begin{abstract}
\Ac{DL} enables collaborative learning without a server and without training data leaving the users' devices.
However, the models shared in DL can still be used to infer training data.
Conventional %
defenses such as differential privacy and secure aggregation fall short in effectively safeguarding user privacy in \ac{DL}, either sacrificing model utility or efficiency.
We introduce \sys, a novel DL approach in which nodes create \acp{VN} to disseminate chunks of their full model on their behalf.
This enhances privacy by 
\begin{enumerate*}[label=\emph{(\roman*)}]
	\item preventing attackers from collecting full models from other nodes, and 
	\item hiding the identity of the original node that produced a given model chunk.
\end{enumerate*}
We theoretically prove the convergence of \sys and provide a formal analysis demonstrating how \sys reduces the efficacy of attacks compared to when exchanging full models between nodes.
We evaluate the convergence and attack resilience of \sys with existing DL algorithms, with heterogeneous datasets, and against three standard privacy attacks. %
Our evaluation shows that \sys not only renders these privacy attacks infeasible when each node operates 16 VNs but also exhibits a positive impact on model utility compared to standard \ac{DL}.
In summary, \sys enhances the privacy of \ac{DL} while maintaining the utility and efficiency of the model.
\end{abstract}


\maketitle

\section{Introduction}

\begin{figure*}[ht]
    \centering
    \includegraphics[width=.9\textwidth]{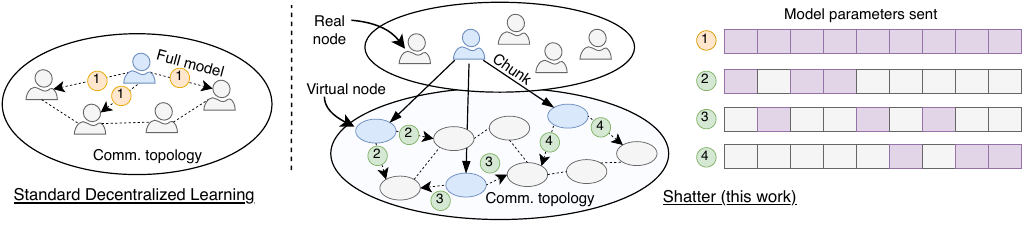}
    \caption{With standard \acf{DL} (left), nodes continuously send their full model to other nodes in the communication topology. With \sys, nodes operate multiple \acp{VN} (middle) and \acp{VN} directly communicate with other \acp{VN}. In addition, each \ac{VN} sends a part of the full model of the \ac{RN} to other \acp{VN} (right). This hides the identity of the original node that produced a given model chunk.} %
    \label{fig:virtual_nodes_architecture}
\end{figure*}

In \Acf{DL}, %
nodes %
collaboratively train a global \ac{ML} model without sharing their private data with other entities~\cite{lian2017can}.
The nodes %
are connected to other nodes, called \emph{neighbors}, via a communication topology.
In each round, nodes start with their local models and perform training steps with their private datasets.
Updated local models are exchanged with neighbors in the communication graph and aggregated.
The aggregated model is then used %
to start the next round, and the process repeats until %
convergence. 
Popular \DL algorithms include \Ac{D-PSGD}~\cite{lian2017can}, \Ac{GL}~\cite{ormandi2013gossip}, and \Ac{EL}~\cite{devos2023epidemic}.

In \ac{DL}, the private data of participating nodes never gets shared. %
While this is a major argument in favor of the privacy of \DL solutions, recent works have shown that model updates can be prone to privacy violations in \DL~\cite{cyffersMuffliatoPeertoPeerPrivacy2022,pasquini2023security}.
For example,
\begin{enumerate*}[label=\emph{(\roman*)}]
\item \acp{MIA}~\cite{shokriMembershipInferenceAttacks2017} allow attackers to infer if a particular data sample was used during training by a node, 
and 
\item \acp{GIA}~\cite{geipingInvertingGradientsHow2020, zhu2019deep} allow the attackers to reconstruct private data samples used for training. %
\end{enumerate*}
These privacy breaches deter the adoption of \ac{DL} algorithms in %
domains where privacy is crucial, like healthcare and finance.
Although numerous strategies to protect the users' privacy have been proposed in the centralized settings for \ac{ML}~\cite{li2021survey}, there is a lack of adequate defenses against privacy attacks in \ac{DL}.
For instance, noise-based solutions bring privacy at the cost of convergence~\cite{yue2023gradient, carlini:2022:mia1stprinciples}, and secure computation schemes provide privacy at the cost of intricate node coordination~\cite{bonawitz2017practical, xu2023secure, lu2023privacy}, resource overhead~\cite{xu2023secure}, or specialized hardware~\cite{cheng2021separation,huba2022papaya}.

To address the privacy concerns of \DL without compromising its utility and efficiency, we introduce \sys, a novel \ac{DL} system that protects shared model updates from these privacy attacks.
\sys consists of three key components: \textit{chunking}, \textit{full sharing}, and \textit{virtualization}.
\textit{Chunking} limits the access of receiving nodes to a model chunk, \ie, a subset of all parameters rather than the full model, thus contributing to privacy.
\textit{Full sharing} ensures that there is no information loss as the sender shares each model parameter with multiple nodes through one of the chunks, thus ensuring model utility.
\textit{Virtualization} decouples identities from model chunks by having each node operate multiple \acfp{VN} that communicate model chunks with other \acp{VN}, thus further contributing to privacy.
Additionally, we randomize the communication topology in each round to prevent an adversary from structurally attacking a fixed set of nodes and to boost model utility.
Compared to standard \ac{DL}, our approach does not sacrifice efficiency and only comes with a manageable increase in communication volume.
Chunking may resemble sparsification, a communication-efficient technique where nodes share subsets of their full model with others~\cite{alistarh2018sparseconvergence}.
However, the \emph{full sharing} of \sys disseminates \emph{all} the model chunks to different \acp{VN}, ensuring that all model parameters of each node are shared and aggregated in each communication round.
Sparsification does not guarantee that all model parameters are sent to other nodes.
While studies have suggested that sharing small model chunks can potentially be privacy-friendly %
as less information is shared among participants~\cite{wei2020framework, shokri2015privacy,zhu2019deep}, to the best of our knowledge, we are the first to leverage this to defend against privacy-invasive attacks in \ac{DL}. %

We visualize the overall architecture of \sys in~\Cref{fig:virtual_nodes_architecture}.
Standard \ac{DL} algorithms (left) connect nodes directly in a communication topology, and nodes exchange their full model with neighbors every round.
In \sys (right), each node, which we refer to as a \acf{RN} in the context of \sys, starts by creating some \acfp{VN} for itself.
All the \acp{VN} then participate in the communication topology construction.
In \sys, \acp{RN} perform the training and chunking of the models, whereas \acp{VN} are responsible for disseminating model chunks through the communication topology.
The right part of~\Cref{fig:virtual_nodes_architecture} highlights how different \acp{VN} send different model chunks, where grayed-out chunks are not sent.
Finally, model chunks received by \acp{VN} are forwarded back to their corresponding \ac{RN} and aggregated there.
The next round is then initiated, which repeats until model convergence.

We formally prove the convergence of \sys and theoretically demonstrate its privacy guarantees that provide an information-theoretical interpretation of the experimental results, illustrating the comprehensive privacy-preserving properties of \sys.
We implement \sys to empirically evaluate its robustness against \emph{honest-but-curious} adversaries that mount three standard attacks: the linkability, membership inference, and gradient inversion attack.
Our experimental results show that \sys provides significantly better privacy protection than baseline approaches while improving convergence speed and attainable model accuracy, at the cost of a manageable increase in communication costs.

In summary, we make the following contributions:
\begin{itemize}
    \item We introduce \sys, a novel privacy-preserving \ac{DL} algorithm where \acfp{RN} operate multiple \acfp{VN} and \acp{VN} share model chunks with each other (Sections~\ref{sec:motivation} and~\ref{sec:design}).
    \item We prove the convergence of \sys with an arbitrary number of local training steps between each communication round. Our bounds involve regularity properties of local functions, the number of local steps, the number of \acp{VN} operated by each \ac{RN}, and the degree of random graphs sampled at each communication round
    (Section~\ref{sec:convergence_analysis}).
    \item We formally show that \sys improves the privacy of \acp{RN} from an information-theoretical perspective as the number of \acp{VN} operated by each \ac{RN} increases. This offers analytical insight into the diminishing efficacy of attacks exploiting shared model parameters or gradient updates (Section~\ref{sec:privacy_analysis}).
    \item We implement \sys and empirically compare its privacy robustness against state-of-the-art baselines and three privacy-invasive attacks: the \ac{LA}, \ac{MIA} and \ac{GIA} (\Cref{sec:evaluation}).
    Our results show that \sys improves model convergence and exhibits superior privacy defense against the evaluated privacy attacks while also showcasing higher model utility. %
\end{itemize}

\section{Background and Preliminaries}
\label{sec:prelims}
In this work, we consider a decentralized setting where a set of nodes $\mathcal{N}$ seek to collaboratively learn an \ac{ML} model.
This is also known as \acf{CML}~\cite{soykan2022survey,pasquini2023security}.
Each node $i \in \mathcal{N} $ has its private dataset $D_i$ to compute local model updates. The data of each node never leaves the device.
The goal of the training process is to find the parameters of the model $\theta$ that perform well on the union of the local datasets %
by minimizing the average loss %
across all the nodes in the network.
The most adopted \ac{CML} approach is \acf{FL} which uses a parameter server to coordinate the learning process~\cite{mcmahan2017communication}.
\Acf{DL}~\cite{nedic2016stochastic,lian2017can,assran2019stochastic} is \iac{CML} algorithm in which each node exchanges model updates with its neighbors through a communication topology comprising of an undirected graph $\cG=(\mathcal{N},\cE)$ where $\mathcal{N}$ denotes the set of all nodes and $(i,j) \in \cE$ denotes an edge or communication channel between nodes $i$ and $j$ (\Cref{fig:virtual_nodes_architecture}, left).
Among many variants, \dpsgd~\cite{lian2017can,pmlr-v119-koloskova20a} is considered a standard algorithm to solve the \DL training tasks.
At the start of \dpsgd, each node $ i $ , with  its loss function $f_i$, identically initializes its local model $\theta_i^{(0)}$ and executes the following: 

\begin{enumerate}[leftmargin=0.4cm]
\item \emph{Local training.} In each round $0\leq t \leq T-1$ and each epoch $0\leq h \leq H-1$, setting $\tilde \theta_i^{(t,0)}$ as $\theta_i^{(t)}$, 
node $i$ independently takes samples $\xi_i$ from its local dataset, computes the stochastic gradient $\nabla f_i(\tilde \theta_i^{(t,h)}, \xi_i) $, and updates its local model as $\tilde\theta_i^{(t,h+1)} \leftarrow \tilde\theta_i^{(t,h)} - \eta \nabla f_i(\tilde\theta_i^{(t,h)}, \xi_i)$, where $ \eta $ is the learning rate. 
\item \emph{Model exchange.} Node $i $ sends $\tilde\theta_i^{(t,H)}$ to and receives $\tilde\theta_j^{(t,H)} $ from each of its neighbors $j$ in $\cG$.
\item \emph{Model aggregation.} Node $i$ mixes the local model parameters that it receives from its neighbors with its own using a weighted aggregation scheme as
$\theta_i^{(t+1)} = \sum_{\{j: (i,j)\in \cE \} \cup \{i\}} w_{ji} \tilde\theta_j^{\left(t,H\right)}, \quad $ where $ w_{ji} $ is the $(j,i)^{\operatorname{th}}$ entry of the mixing matrix $ W $. A common approach is to %
aggregate all %
the models %
with equal weights.
\end{enumerate}

After $ T $ rounds, node $ i $ adopts the final $ \theta_i^{(T)} $, 
thereby terminating the \ac{DL} training.
The \dpsgd pseudocode is also provided in \Cref{alg:dpsgd} in \Cref{app:dpsgd}.

\subsection{Privacy Attacks in \ac{CML}}
\label{sec:prelims_attacks}
A key property of \ac{CML} algorithms is that training data never leaves the node's device, therefore providing some form of data privacy.
Nonetheless, an adversarial server or node may be able to extract sensitive information from the model updates being shared with them.
We next outline three prominent privacy attacks in \ac{CML}.

\paragraph{\textbf{Membership inference attack (MIA)}}
The goal of the \ac{MIA} is to correctly decide whether a particular sample has been part of the training set of a particular node~\cite{hu2022membership}.
This is broadly a black-box attack on the model, with the adversary having access to the global training set and samples from the test set, which are never part of the training set of any node.
While we assume that the adversary can access actual samples from the global training and test set, this can be relaxed by generating shadow datasets~\cite{shokriMembershipInferenceAttacks2017}.
Many \ac{MIA} attacks are based on the observation that samples included in model training exhibit relatively low loss values as opposed to samples that the model has never seen.
Since the \ac{MIA} can present a major privacy breach in domains where sensitive information is used for model training, this attack is widely considered a standard benchmark to audit the privacy of \ac{ML} models~\cite{ye2022enhanced}.

We evaluate the privacy guarantees of \sys using a loss-based \ac{MIA}, mainly because of its simplicity, generality, and effectiveness~\cite{pasquini2023security}.
The adversary queries the received model to obtain the loss values on the data samples.
The negative of the loss values can be considered as \emph{confidence scores}, where the adversary can use a threshold to output \ac{MIA} predictions~\cite{yeom2018privacy}.
To quantify the attack regardless of the threshold, it is common to use the ROC-AUC metric (\acl{AUC}) on the confidence scores~\cite{liu2022ml,pyrgelis2017knock}. 
Specifically, membership prediction is positive if the confidence score exceeds a threshold value $\tau$ and negative otherwise.
Hence, we can compute the corresponding \ac{TPR} and \ac{FPR} for each $\tau$, resulting in the \ac{ROC} curve.

\paragraph{\textbf{Gradient inversion attack (GIA)}}
With the \Ac{GIA}, the adversary aims to reconstruct input data samples from the gradients exchanged during the training process in \ac{CML}~\cite{zhu2019deep,geipingInvertingGradientsHow2020}.
This is a key privacy violation in contexts involving sensitive information, such as personal photographs, medical records, or financial information.
Since the success of this attack depends on the information contained in the gradients, this white-box attack in \ac{DL} is performed at the first communication round or convergence, when the gradient approximation by the adversary is the best~\cite{pasquini2023security}.
The \Ac{GIA} is an optimization problem where the adversary performs iterative gradient descent to find the input data that results in the input gradients~\cite{zhu2019deep, deng2021tag}.
More sophisticated \acp{GIA} on image data include \textsc{GradInversion}~\cite{yin2021see} when the training is done on a batch of images, and \ROG~\cite{yue2023gradient} using a fraction of gradients~\cite{yue2023gradient}.
We use the \textit{state-of-the-art} \ac{GIA} scheme \ROG to evaluate \sys.
In addition, we leverage artifacts and pre-trained weights of reconstruction networks provided with \ROG and assume that the adversary knows the ground-truth labels of samples in the batch.
These, however, can also be analytically obtained from the gradients of the final neural network layer~\cite{zhao2020idlg, wainakh2021label}.

\paragraph{\textbf{Linkability attack (LA)}}
In the context of obfuscated model updates, the \ac{LA} allows an adversary to link a received obfuscated update with the training set it came from~\cite{lebrun2022mixnn}.
Similar to the \ac{MIA}, this is a loss-based black-box attack on the model, but contrary to \ac{MIA}, the adversary has access to the training sets of each participating node.
The \ac{LA} may also be performed on shadow data representing the training sets of participating nodes instead of actual data.
In the context of this work, we assume that the adversary can access the actual training sets.
An adversary performs a \ac{LA} on received model updates by computing the loss of each received model update on each available training set and reporting the training set with the lowest loss.
When the obfuscated update does not contain the complete model update vector, \eg, when using model sparsification, the adversary completes the vector with the aggregated model updates from the previous round~\cite{lebrun2022mixnn}.

\subsection{Threat Model}
\label{sec:threatModel}

This work focuses on privacy-preserving \ac{DL} in a \emph{permissioned network setting} where membership is strictly controlled and well-defined.
This aligns with the observation that \ac{DL} is commonly considered and deployed in enterprise settings, where network membership is typically controlled~\cite{beltran2023decentralized}.
Such settings include, for example, hospitals collaborating on a \ac{DL} task~\cite{shiranthika2023decentralized}.
In permissioned networks, all nodes are known entities with verified identities.
Consequently, we consider threats commonly found in open networks, such as the Sybil Attack~\cite{douceur2002sybil}, beyond the scope of our work.

Our work focuses on the \emph{honest-but-curious} (HbC) attacker model, \ie participating nodes faithfully execute the learning protocol but can attempt to retrieve sensitive information about the other nodes from locally-available information.
This attacker model is commonly adopted in related work on privacy in \ac{CML} algorithms~\cite{cyffersMuffliatoPeertoPeerPrivacy2022, pmlr-v151-cyffers22a, lu2023privacy, xu2023secure, yue2023gradient}.
Thus, we adhere to a threat model that is locally privacy-invasive, allowing any update that a node shares with any other node to be potentially exploited to infer information about their personal data used for training their local model in each round.
In particular, we do not consider that aggregated models of any node in any round of communication are published or shared with any third-party entity (\eg a server). 
We also limit ourselves to a setting where only participating nodes can be adversaries attempting to compromise the privacy of other nodes in the network.
Besides received model updates, attacker nodes may also use knowledge of their own model parameters to carry out attacks, and may store historical model updates or other information they received and act on this.
Furthermore, all nodes can potentially be attackers. However, we assume that nodes do not collude between them and do not maliciously modify the model parameters.
Such local-level privacy risks (\ie any information that leaves a node can be used against them) are widespread in the literature~\cite{NISTThreatModels2020} and in line with some of the recently popularized threat models considered in practice~\cite{GoogleRappor2014, AppleDP2017} that do not assume the presence of any trusted entity responsible for aggregation or orchestrating the communication. 
\begin{figure}[t!]
	\centering
	\includegraphics[width=\linewidth]{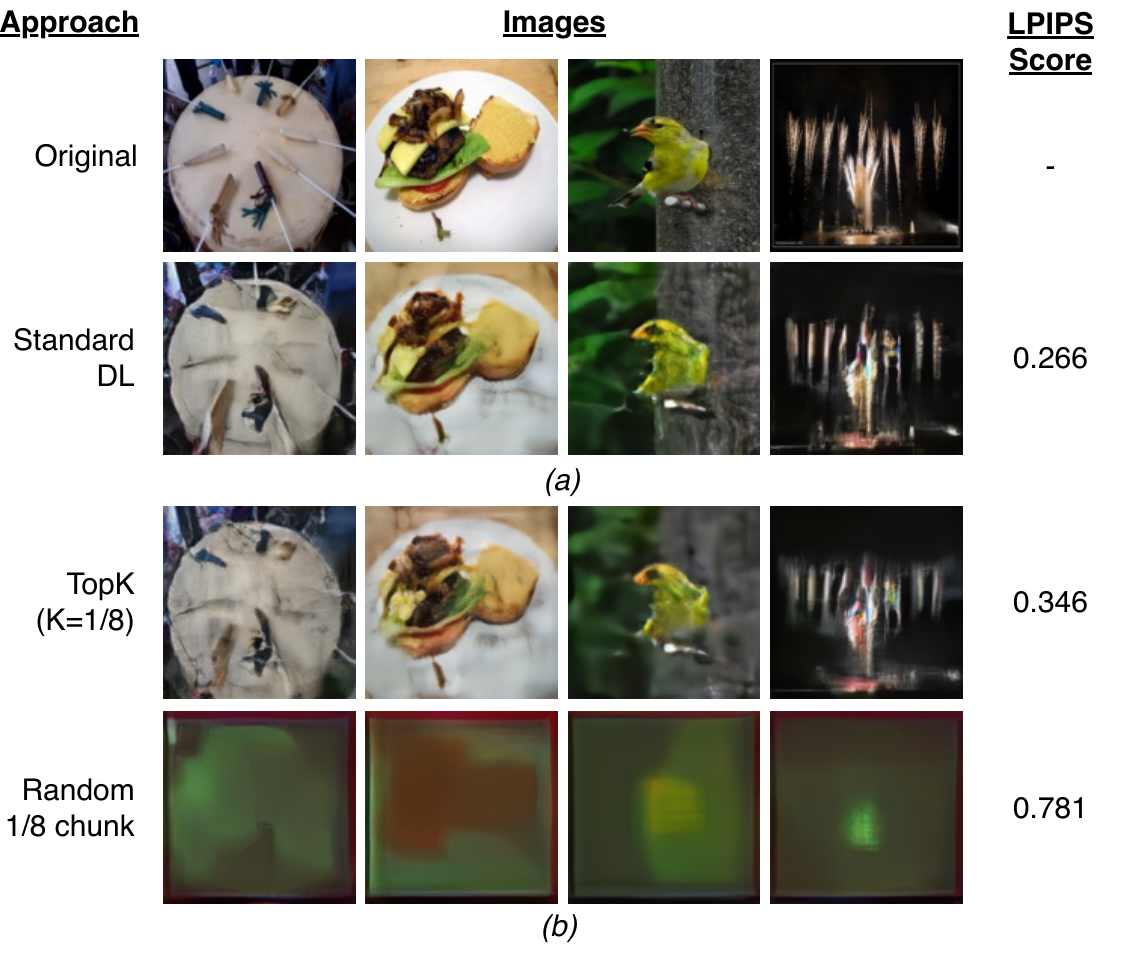}
	\caption{Selected reconstructed images (per row) using the \ac{GIA} and the average LPIPS score ($\uparrow$ is more private) for all 1600 images processed in a round, when using standard 
 \ac{DL} with 100 clients (a), \topk sparsification and random model chunking (b).}
	\label{fig:gradient_inversion_images_motivation}
\end{figure}

\section{Motivation behind \sys}
\label{sec:motivation}
The design of \sys is anchored in three key insights into the privacy challenges faced by conventional \ac{DL} systems.

\label{sec:motivation_privacy}
\paragraph{\textbf{1) Naive (full) model sharing in \DL reveals sensitive data}}
A key aspect of collaborative \ac{ML} mechanisms is that private data never leaves the devices of participating users.
While this might give a sense of security, Pasquini~\etal~\cite{pasquini2023security} recently demonstrated how model sharing in standard \ac{DL} settings can reveal sensitive information.
To show this empirically, we make each node conduct a \ac{GIA} during the first training round on the incoming models in a standard \ac{DL} setting with \num{100} nodes.
Conducting the \ac{GIA} during the first training round (or close to convergence) is optimal for the attack's success~\cite{pasquini2023security}.
Each node trains its local model (LeNet) using a batch with \num{16} images from ImageNet, and adversaries (all other nodes) attempt to reconstruct the images in this batch using \ROG~\cite{deng2009imagenet}.
We further elaborate the experiment setup in~\Cref{subsec:exp_setup}.

The results of \ROG in \DL for four random images are shown in~\Cref{fig:gradient_inversion_images_motivation}a, with the original image in the top row and the reconstructed image by the attacker in the second row.
For each approach, we also show the average LPIPS score of all \num{1600} images processed during a training round~\cite{zhang2018unreasonable}.
This score indicates the perceptual image patch similarity, where higher scores indicate more differences between the original and reconstructed image, \ie, the higher the more private.
We observe significant similarities between the original and reconstructed images in \ac{DL} settings, making it trivial for an attacker to obtain knowledge and semantics of private training data of other users.
Empirical findings highlighted by other \ac{DL} studies~\cite{pasquini2023security} and our experiments illustrate that \emph{naive model sharing in \ac{DL} falls short as a method for safeguarding data privacy}, as they reveal sensitive data even in early training rounds.

\paragraph{\textbf{2) Partial model sharing can protect privacy}}
\label{sec:motivation_sparsification}
Intuitively, sending only a subset of model parameters to other nodes raises the bar for adversaries to obtain sensitive information, and can be leveraged to increase privacy in \ac{DL} systems.
We note, however, that partial model sharing, also known as sparsification, is typically employed to reduce communication cost~\cite{shokri2015privacy, strom2015scalable, alistarh2018sparseconvergence, lin2018deep,koloskova2019decentralized, koloskova2020decentralized}.
Instead of sharing all model parameters, with sparsification, nodes only send a fraction of the model updates, \ie, each node $ i $ constructs and shares a sparsified model $ S_i^{(t)} \subseteq \theta_i^{(t)} $ in round $ t $.
Typical sparsification approaches retain random parameters (random sharing) or the ones with the highest gradient magnitudes (\topk sharing).

While the work of Yue~\etal demonstrates how \topk sparsification gives a false sense of privacy~\cite{yue2023gradient}, there are key differences in privacy guarantees between standard sparsification techniques.
To understand the privacy implications of using \topk and random sharing, we conduct the \ac{GIA} with these sharing methods.
\Cref{fig:gradient_inversion_images_motivation}b shows the reconstructed images with \ROG when the network uses \topk sharing with $ \frac{1}{8} $th of the model updates being communicated.
Even though the LPIPS score related to using \topk sharing is higher than sharing full models (second row in~\Cref{fig:gradient_inversion_images_motivation}a), it is evident that \topk sharing still allows an attacker to reconstruct potentially sensitive information from private training datasets.
This is because \topk shares the parameters with the highest gradient magnitudes, and hence, the ones that are the most influential in the training round.
Access to these parameters raises the effectiveness of the \ac{GIA} that relies on gradients to reconstruct private data.

The bottom row in~\Cref{fig:gradient_inversion_images_motivation}b shows the reconstructed images by an adversary when each node randomly sends $ \frac{1}{8} $th of its model update to neighboring nodes (random sharing).
Random sharing yields blurry images, making it nearly impossible for an adversary to obtain semantic information.
This is also demonstrated by the associated LPIPS score, which is nearly three times as high as when using \topk sharing.
While it is evident that \emph{random sharing is effective at obfuscating updates in \ac{DL}}, partial model sharing has been shown to hurt time-to-convergence and final achieved accuracy, therefore prolonging the overall training process~\cite{hsiehskewscout2020,dhasade2023get,dhasade2023decentralized}.%

\begin{figure}[b!]
	\centering
	\inputplot{plots/motivation}{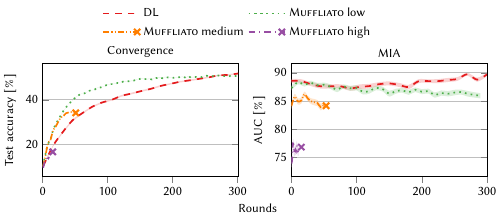}
    	\caption{The test accuracy ($\uparrow$ is better) on the left and \ac{MIA} attack success ($\downarrow$ is more private) on the right for \ac{DL} and \muffliato with three noise levels, using \cifar
     as training dataset.}%
	\label{fig:motivation_noise}
\end{figure}

\begin{figure*}[t!]
    \centering
    \includegraphics[width=\textwidth]{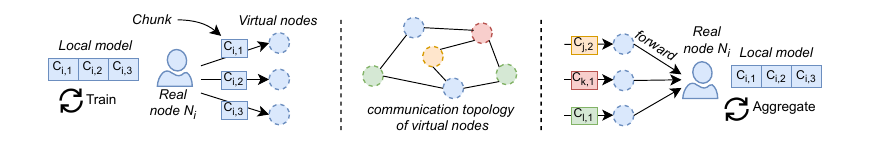}
    \caption{The three-step workflow of \sys, showing the operations during a single training round. A \ac{RN} $ N_i $ first splits its local model into $ k $ chunks and sends each chunk to one of its \acp{VN} (left). We refer to the $s^{th}$ chunk of \ac{RN} $ N_i $ as $ C_{i,s}$. \Acp{VN} are connected into a communication topology and exchange model chunks with other \acp{VN} (middle). \Acp{VN} forward received chunks to the corresponding \ac{RN} who aggregates the received chunks into its local model and performs a training step (right). }
    \label{fig:virtual_nodes_algorithm}
\end{figure*}

\paragraph{\textbf{3) Noise-based solutions have drawbacks}}
\label{sec:motivation_noise}
Another approach to increase privacy in \ac{CML} is to add small perturbations (\textit{a.k.a.} noise) to model updates before sharing them with other nodes~\cite{wei2021user, yu2021not}.
Noise-based mechanisms for \ac{DL} with differential-privacy foundations like \muffliato~\cite{cyffersMuffliatoPeertoPeerPrivacy2022} have been theoretically proven to converge while providing strong privacy guarantees.
In practice, though, adding noise for privacy protection can hurt the utility of the model~\cite{yue2023gradient, pasquini2023security}.
We illustrate this by evaluating the convergence and the success of the \ac{MIA} for both \EL, a \DL variant where the communication topology is refreshed every round, and \muffliato on a \niid partitioning of the \cifar dataset with a \resnet model.
We comprehensively search for three noise levels in \muffliato, one with an attack success close to \DL, one with a low attack success rate, and one in between.
We run \muffliato with \num{10} communication rounds as recommended by the authors of \muffliato~\cite{cyffersMuffliatoPeertoPeerPrivacy2022}.
While this increases the communication cost of \muffliato by $10\times$ compared to \DL, we consider this fair from a convergence and privacy perspective.

\Cref{fig:motivation_noise} shows our results and demonstrates the pitfalls of the \textit{state-of-the-art} noise-based \DL solution \muffliato.
The figure shows the test accuracy and \ac{MIA} attack resilience for \DL and \muffliato for three different noise levels (low: $ \sigma = 0.025$, medium: $ \sigma = 0.05$, high: $ \sigma = 0.1 $).
While \muffliato~(\textit{low}) does reduce the \ac{MIA} success rate when compared to \DL, this reduction is marginal.
The better convergence of \muffliato~(\textit{low}) can be attributed to the \num{10} communication rounds in \muffliato, as opposed to \num{1} in \DL. 
Extra communication rounds enable nodes to better aggregate the models across the network.
We also observe that model training with \muffliato~(\textit{medium}) breaks down after roughly \num{45} rounds into training.
This is because noise introduces numerical instability in the learning process, leading to \textit{nan} (not a number) loss values.
On the other hand, \muffliato~(\textit{high}) handicaps the model utility, as there is no convergence (\Cref{fig:motivation_noise}, left). 
Its better performance in \ac{MIA} is, hence, not representative of the learning process.
We emphasize that there is no way to predict the noise levels in \muffliato, and in similar solutions. %
Furthermore, Yue~\etal empirically show that noise-based solutions hurts the final accuracy when reducing the success of \ac{GIA}~\cite{yue2023gradient}.
The results in~\Cref{fig:motivation_noise} highlight the \emph{need for practical privacy-preserving solutions that do not adversely affect the convergence of \DL tasks}.

\section{Design of \sys}

\label{sec:design}
Based on the observations in~\Cref{sec:motivation}, we present the design of \sys.
The essence of \sys is to exchange a randomly selected subset of parameters rather than the entire set (\textit{chunking}) while still disseminating all the parameters through the network to not lose convergence (\textit{full sharing}).
\sys achieves this by having each \acf{RN} operate multiple \acfp{VN} (\textit{virtualization}) and the \acp{VN} communicate random model chunks with other \acp{VN} in the network.
Our design enhances the privacy of \ac{DL} as it significantly complicates the task of adversarial \acp{RN} to extract sensitive information from received model chunks.

\subsection{System Model}
\label{sec:prelims_threat}

In \sys, each participating \ac{RN} spawns and operates \acp{VN}.
We call an \ac{RN} $N_i$ the \emph{parent} of a certain \ac{VN} if the latter was spawned by $N_i$.
Each \ac{VN} has a unique identifier that identifies the node in the network.
We assume a permissioned network, a standard assumption of many \ac{DL} approaches.
In the $t^{\operatorname{th}}$ communication round, let $\cG_t=(\cV,\cE_t)$ be the graph on the \acp{VN}, where $\cV$ is the set of \acp{VN} 
and $\cE_t$ is the set of edges (communication) established between the \acp{VN} in round $t$.
\sys connects \acp{VN} using a $r$-regular communication graph, which is a commonly used topology across \ac{DL} algorithms due to its simplicity and load balancing.
With this topology, each \ac{VN} has exactly $ r $ incoming and outgoing edges, \ie, each \ac{VN} sends to and receives from $ r $ other \acp{VN}.
The convergence of \ac{DL} on $r$-regular graphs has extensively been studied~\cite{devos2023epidemic}.
\acp{RN} and \acp{VN} can go offline or come back online during model training.

In line with~\Cref{sec:threatModel}, we consider all \acp{RN} to be honest-but-curious, \ie the participating \acp{RN} faithfully execute the protocol but may attempt to retrieve sensitive information about the other \acp{RN} from the model updates it receives through its \acp{VN}.
We assume that every \ac{VN} works in the best interest of their corresponding \acp{RN}, \ie, an adversarial \ac{RN} can leverage all its \acp{VN} to participate in the attack.
We consider that the \acp{RN} have full control over their respective \acp{VN} in a sense that: $i)$ every \ac{RN} knows which other \acp{VN} its own \acp{VN} have communicated with in a given training round, and $ii)$ the information about which parent \ac{RN} controls a given \ac{VN} cannot be retrieved by any external \ac{VN} or \ac{RN} (\ie the link between any \ac{RN} and its \acp{VN} is hidden from everyone else in the network). %

\subsection{\sys Workflow}
\label{sec:vnodes}
We show the full workflow of \sys in~\Cref{fig:virtual_nodes_algorithm} and provide pseudocode in~\Cref{alg:vnodes}. The key notations used to describe and analyze the working of \sys are summarized in \Cref{app:notations}.

In \sys, each participating \acf{RN} creates a set of $ k $ \acp{VN} for itself (Line~\ref{line:spawn_vns}).
These \acp{VN} are, in practice, implemented as processes on remote systems.
To streamline our analysis, we work under the assumption that each \ac{RN} operates the same number of \acp{VN}.
We show in~\Cref{sec:privacy_analysis} that increasing $ k $, \ie, spawning more \acp{VN}, reduces the vulnerability against privacy attacks.

At the start of each training round $t$, each \ac{RN} first updates its local model using its private dataset (Lines~\ref{line:vnodes_local_update_start}--\ref{line:vnodes_local_update_end}).
The \ac{RN} then segments its local model into $ k $ smaller model chunks (\emph{chunking}-specifications are given later) and forwards these to its \acp{VN} %
that multicast the model on the \ac{RN}'s behalf (\emph{full sharing} -- Lines~\ref{line:vnodes_chunking_start}--\ref{line:vnodes_chunking_end}).
We also visualize this in~\Cref{fig:virtual_nodes_algorithm} (left) where some real node $ N_i $ segments its local model into three chunks $ C_{i,1}, C_{i,2} $, and $ C_{i,3} $, each of which is forwarded to a \ac{VN} of $ N_i $.
In standard \ac{DL} algorithms, \acp{RN} directly communicate their model updates.
\sys instead connects \acp{VN} in a communication topology $ \cG_t$ that is randomized every round (\Cref{line:randomize_topology}), also see~\Cref{fig:virtual_nodes_algorithm} (middle).
In \sys, \acp{VN} thus act as communication proxy for model chunks created by \acp{RN}.
All model chunks received by a \ac{VN} during a training round are forwarded back to its parent \ac{RN} and aggregated into the local model (\Cref{line:vnodes_aggregation}).
This is also visualized in~\Cref{fig:virtual_nodes_algorithm} (right) where an \ac{RN} $ N_i $ receives model chunks from the \acp{VN} operated by \acp{RN} $ N_j, N_k $, and $ N_l $.
We note that \ac{RN} $ N_i $ is oblivious to the identity of the \acp{RN} behind the received model chunks.
In essence, from the perspective of a single real node, it sends out the same model parameters as in standard \DL.
While a real node sends a smaller subset of random parameters to one virtual node, it also sends the remaining parameters in chunks to other virtual nodes, which ultimately enhances privacy without adverse effects on convergence.

\begin{algorithm2e}[t]
    \DontPrintSemicolon
    \caption{\sys from the perspective of \ac{RN} $ N_i $}
    \label{alg:vnodes}
    Initialize $\theta_i^{(0)}$\;
    Spawn $ k $ \acp{VN}: $v_i(1),\ldots v_i(k)$\label{line:spawn_vns} \;
    \For{$t = 0, \dots, T-1$}{
    $\tilde \theta_i^{(t,0)}\leftarrow\theta_i^{(t)}$\;
     \For{$h = 1, \dots, H$}{\label{line:vnodes_local_update_start}
            $\xi_i \leftarrow$ mini-batch sampled from $D_i$ \;
            $\tilde\theta_i^{(t,h+1)} \leftarrow \tilde\theta_i^{(t,h)} - \eta \nabla f_i(\tilde\theta_i^{(t,h)}, \xi_i)$\;
        }\label{line:vnodes_local_update_end}
    \textbf{Chunk} $\tilde\theta_i^{(t,H)}$ into $ k $ chunks\label{line:vnodes_chunking_start}\;
    \For{$s=1,\dots,k$}
    {\textbf{Forward} chunk $ s $ to $v_i(s)$ for every $s\in[k]$\;%
    \label{line:vnodes_chunking_end}}
    Randomize communication topology $ \cG_t$ \label{line:randomize_topology} \;
    \textbf{Receive} $r$ chunks from each of the $k$ \acp{VN}\;
    \textbf{Aggregate} the received chunks to produce $\theta_i^{(t+1)}$\label{line:vnodes_aggregation}
    }
    \KwRet $\theta_i^{(T)}$\;
\end{algorithm2e}

\paragraph{\textbf{Model chunking strategy.}}
Each node randomly samples parameters to chunk its local model without replacement.
With this strategy, a \ac{RN} sends distinct parameters to its \acp{VN}.
Thus, in one round, two chunks from the same \ac{RN} will always be disjoint.
and, hence, an adversary cannot put together two chunks originating from the same real node based on an intersection of parameters.

Additionally, we use \emph{static} model chunking (\ie, fixed across rounds) where all \acp{RN} follow an identical chunking strategy (\ie, they choose the exact same partitions by having a shared seed).
That is, each \ac{VN} %
is responsible for the same set of parameter indices across rounds.
We acknowledge that alternative model chunking strategies are possible, and describe some of these strategies and their impact on privacy in~\Cref{app:alternative_chunking_strategies}.

\paragraph{\textbf{Dynamic topologies}}
To enhance the privacy and efficiency of \sys, we build \sys over \ac{EL}, the \textit{state-of-the-art} \ac{DL} algorithm where the communication topology $ \cG_t$ is refreshed in every training round $t$.
The benefits of having dynamic topologies over static topologies in \sys are twofold.
Firstly, it restricts any attacker from receiving model chunks consistently from a \ac{VN} of the same victim node.
This reduces the chances of attacks that target specific nodes as an attacker is unable to reliably obtain model chunks from the victim across training rounds.
Secondly, dynamic topologies converge faster than static ones thanks to the better mixing of the models~\cite{devos2023epidemic}.
We argue, however, that the components of \sys are generic enough to be compatible with other \DL baselines, such as \dpsgd or variants.
Refreshing $ \cG_t$ each round can be achieved by having \acp{VN} participate in topology construction using a decentralized peer-sampling service~\cite{voulgaris2005cyclon,jelasity2007gossip,nedelec2018adaptive}.
In this paper, we stick to the \emph{EL-Oracle} model~\cite{devos2023epidemic} in which a central coordinator agnostic of the identities of \acp{RN} creates a random $r$-regular topology of only \acp{VN} every training round.
We assume that nodes faithfully participate in the construction of the communication topology, which can be achieved by using accountable peer sampling services~\cite{yoon2023accountnet,antonov:2023:securecyclon}.

\paragraph{\textbf{Model aggregation}}
After the \acp{VN} forward the received model parameters back to their parent \ac{RN}, each \ac{RN} receives the same total number of parameters in a \emph{r}-regular topology, but the count of incoming parameters at a particular index may differ.
\sys uses a parameter-wise weighted averaging where the weight is proportional to the frequency of how often this parameter is received.
For example, if a parameter $ p $ is received two times, $ p_1 $ and $ p_2 $ referring to these individual parameters, a \ac{RN} will average $ p_1 $, $ p_2 $ and the same parameter in its local model while using an averaging weight of $ \frac{1}{3} $ for each parameter.
Moreover, it might occur that a \ac{RN} does not receive a particular parameter at all, in which case it will simply adopt the parameter in its local model.
\section{Theoretical Analysis}\label{sec:analysis}
%

%

%
\paragraph{Notations.}

Let $\mathcal{N}=\{N_1,\ldots,N_n\}$ be the \acp{RN} in the network. Setting $d\in\mathbb{N}$ as the dimension of the parameter space, let $\theta^{(t)}_i=(\theta^{(t)}_{i}(1),\ldots,\theta^{(t)}_{i}(d))\in\mathbb{R}^d$ be the model held by $N_i$ in round $t$ for every $i\in[n]$. Each \ac{RN} is assumed to have $k$ \acp{VN}, with $v_i(s)$ denoting the $s^{\operatorname{th}}$ \ac{VN} of $N_i$ for all $i\in[n]$ and $s\in[k]$, and $\cV=\{v_i(s),i\in[n],s\in[k]\}$ being the set of all \acp{VN}. Let $\theta_{i,s}^{(t)}$ be the chunk of $N_i$'s model that is forwarded to $v_i(s)$ in round $t$. 
For every $i,j\in[n]$ and $p\in[d]$, 
let $\theta_{i\leftarrow j}^{(t)}(p)$ denote the $p^{\operatorname{th}}$ parameter of $N_j$'s model that is shared with $N_i$ and, correspondingly, let $\theta^{(t)}_{i\leftarrow j}%
$ be the entire contribution that $N_i$ receives from $N_j$'s model updates (via the interaction of their \acp{VN}) for aggregation leading up to the next communication round. Analogously, let $\theta^{(t)}_{i\not\leftarrow j}$ be the part of $N_j$'s model that $N_i$ has \emph{not} received in round $t$ via their \acp{VN}. Let $\left\lvert \theta^{(t)}_{i\leftarrow j}\right\rvert$ and $\left\lvert \theta^{(t)}_{i\not\leftarrow j}\right\rvert$ denote the number of model parameters of $\theta^{(t)}_j$ that are shared and \emph{not} shared with $N_i$, respectively, in round $t$ (\ie, $\left\lvert \theta^{(t)}_{i\leftarrow j}\right\rvert+\left\lvert \theta^{(t)}_{i\not\leftarrow j}\right\rvert=d$).  In the subsequent analysis, we assume that:

\begin{enumerate}[label=$\roman*.$]
    \item the number of model parameters held by each \ac{VN} is the same (\ie, $d=kc$ for some $c\in\mathbb{N}$). Thus, the chunking partitions the model of the corresponding \ac{RN} into $k$ equal parts, each with $c$ parameters).
    \item the $kn$ total \acp{VN} form an $r$-regular dynamic topology, for some $1\leq r\leq nk-1$, facilitating interaction among them for the exchange of the model chunks they individually hold.
\end{enumerate}

\subsection{Convergence of \sys}\label{sec:convergence_analysis}

For $i\in[n]$, let $f_i:\R^d\to\R$ be the loss of node $N_i$ with respect to its own dataset, and let $f=\frac{1}{n}\sum_{i=1}^n f_i$ the function that the nodes seek to minimize.
Nodes will alternate between rounds of $H\geq 1$ local stochastic gradient steps and communication rounds, that consist of communications through \acp{VN}.

We recall that %
$\cG_t=(\cV,\cE_t)$ is the $r$-regular communication graph on the \acp{VN} in round $t$ %
and let $\cE_t$ (the set of edges of $\cG_t$) be sampled on $\cV$ uniformly at random and independently from the past. %
A communication round consists of a simple gossip averaging in the \ac{VN} graph -- this could be refined with accelerated gossip averaging steps \cite{even2021continuized} or compressed communications \cite{koloskova2019decentralized} for instance. 
Note that for computation steps, \emph{no} %
noise is added, as opposed to other decentralized privacy-preserving approaches, such as \cite{cyffersMuffliatoPeertoPeerPrivacy2022} for gossip averaging or \cite{pmlr-v151-cyffers22a,pmlr-v202-even23a} for token algorithms. As with communications, computations are also assumed to be synchronous, and employing asynchronous local steps, such as asynchronous SGD, would require a more complex and involved analysis~\cite{mishchenko2022asynchronous,pmlr-v70-zheng17b}. We can now write the updates of the training algorithm as follows.
\begin{enumerate}
    \item \textit{Local steps.} 
    For all $i\in[n]$, let $\tilde \theta_i^{(t,0)}=\theta_i^{(t)}$ and for $h\leq H-1$, 
    \begin{equation*}
        \tilde \theta_i^{(t,h+1)}=\tilde \theta_i^{(t,h)}-\eta g_i^{(t,h)} \,,
    \end{equation*}
    where $\eta>0$ is the learning rate and $g_i^{(t,h)}=\nabla f_i (\tilde\theta_i^{(t,h)},\xi_i)$ a stochastic gradient of function $f_i$ computed on the mini-batch $\xi_i$ sampled independently from the past.
    \item \textit{Communication step.} 
    Let $\hat \theta_i^{(t)}=\tilde\theta_i^{(t,H)}$ and for a \ac{VN} $v_i(s)$, let $\hat \theta_{v_i(s)}^{(t)}$ be its model chunk. Then, an averaging operation is performed on the graph of virtual nodes:
    \begin{align*}
        \theta_i^{(t+1)}=\sum_{s=1}^k \frac{1}{r} \sum_{w\in\cV:\{v_i(s),w\}\in\cE_t} \hat \theta_w^{(t)}\,.
    \end{align*}
\end{enumerate}

\begin{theorem}\label{thm:convergence}
    Assume that functions $f_i$ are $L-$smooth, $f$ is lower bounded and minimized at some $\theta^\star\in\R^d$, the stochastic gradients are unbiased (\ie, $\mathbb{E}\left[g_i^{(t,h)}\right] = \nabla f_i \left(\tilde \theta_i^{\left(t,h\right)}\right)\,\forall\,i,t,h)$ of variance upper bounded by $\sigma^2$: $\frac{1}{n}\sum_{i=1}^n\mathbb E \left[\left\| g_i^{(t,h)}-\nabla f_i (\tilde \theta_i^{(t,h)})\right\|^2\right]\leq \sigma^2\,,$
    and let $\zeta^2$ be the population variance, that satisfies:
    \begin{equation*}
        \forall \theta\in\R^d\,,\quad \frac{1}{n}\sum_{i=1}^n \left\| \nabla f_i ( \theta) -\nabla f(\theta)\right\|^2\leq \zeta^2\,.
    \end{equation*}
    Finally, assume that $\rho<1$, where $\rho$ is defined in \Cref{eq:rho} in the proof of \Cref{lem:contraction}.
    Then, for all $T>0$, setting $F_0=f(\bar\theta^{(0)})-f(\theta^\star)$, there exists a constant stepsize $\eta>0 $ such that:
    \begin{align*}
        \frac{1}{T}\sum_{t<T}\mathbb E \left[ \left\| \nabla f(\bar \theta^{(t)})\right\|^2\right]&=\mathcal O\left(\sqrt{\frac{LF_0 \sigma^2}{nHT}} + \frac{LF_0}{T(1-\rho)}\right.\\
        &\quad\left.+\left[\frac{LF_0\big(H\zeta + \sigma\sqrt{(1-\rho)H}\big)}{(1-\rho)HT}  \right]^{\frac{2}{3}}  \right)\,,
    \end{align*}
\end{theorem}

The proof is postponed to \Cref{sec:proof:convergence} and involves an intermediate result given as \Cref{lem:contraction} %
preceding the proof of the main result. 
The convergence bound shows that \sys finds first-order stationary points of $f$ and. As $f$ is non-convex, we fall back to show that the algorithm will find an approximate first-order stationary point of the objective~\cite{carmon2017lower1}.
The first term $\sqrt{\frac{LF_0 \sigma^2}{nHT}}$ is the \emph{statistical rate} and is the largest as $T$ increases. %
$nHT$ is precisely the number of stochastic gradients computed up to iteration $T$, and this term cannot be improved \cite{carmon2017lower1}.
The second and third terms are then lower order terms: for $T=\Omega( n H(1-\rho)^{-2})$, the first term dominates.

Finally, the assumption $\rho<1$ needs to be verified.
For large $n$, it will always be satisfied for $r > 1$. In the $n$-large regime, $\rho$ scales as $\rho\sim \frac{1}{r}$ and, thus, increasing $r$ will always lead to faster communications. However, it appears that $\rho$ only needs to be bounded away from $1$, as only the $1/(1-\rho)$ factor appears in the rate. Having $\rho\leq 1/2$ is sufficient to obtain the best rates, and this only requires $r=\cO(1)$. There is no need to scale $r$ with $n$, and the graph of virtual nodes can have bounded degrees.

\subsection{Privacy Guarantees}\label{sec:privacy_analysis}

Under the assumed threat model (see~\Cref{sec:prelims_threat}) and staying consistent with the experimental evaluation (see~\Cref{sec:exp_privacy_of_nodes}), we have that for any given \ac{RN} $N_j$ in the network, every other \ac{RN} who receives any part of $N_j$'s model in any given round could independently act as a potential HbC adversary trying to compromise $N_j$'s privacy. It is important to recall that, as we do not consider collusion, the privacy-invasive attacks are \emph{pairwise independent} between the \acp{RN}. In other words, if $\mathcal{N}(j,t) \subset \mathcal{N}$ is the set of all the \acp{RN} that receive some parts of $N_j$'s model in a certain round $t$ through the interaction between their corresponding \acp{VN}, we assume that each $N_i \in \mathcal{N}(j,t)$ can \emph{independently} use the information at their disposal to compromise the privacy of $N_j$. However, the neighbors of $N_j$ \emph{do not} exchange information between them to impose a colluded attack against $N_j$. Therefore, in order to examine the privacy guarantees for any \ac{RN} $N_j$, it suffices to carry out the analysis from the perspective of any $N_i \in \mathcal{N}(j,t)$. Hence, in the subsequent analysis, without loss of generality, we fix a certain \ac{RN} $N_j$ as a potential victim and, correspondingly, an HbC adversarial \ac{RN} $N_i \in \mathcal{N}(j,t)$ trying to compromise the privacy of $N_j$, thus proceeding to study how \sys improves the privacy of any \ac{RN} present in the network.

Setting $\theta^{(t)}_{i\leftarrow v_{i}(s)}(p)$ as the model parameter $p\in[d]$ that $N_i$ receives from its \ac{VN} $v_i(s)$ for any $s\in[k]$ after any communication round $t$, for every $j,j'\in [n],\,j\neq j'$, we have:
\begin{align}
    \mathbb{P}\left[\theta^{(t)}_{i\leftarrow v_{i}(s)}(p)=\theta^{(t)}_{i\leftarrow j}(p)\right]=\mathbb{P}\left[\theta^{(t)}_{i\leftarrow v_{i}(s)}(p)=\theta^{(t)}_{i\leftarrow j'}(p)\right].\nonumber
\end{align}
This is because the system model of \sys (as described in \Cref{sec:prelims_threat}) assumes that the \acp{VN} that communicate with $v_i(s)$ do not reveal any information about their respective parent \acp{RN} and, thus, the model parameters transmitted by them to $v_i(s)$ (which, in turn, $N_i$ receives) are equiprobable to come from any of the other participating \acp{RN}. Hence, by achieving perfect indistinguishability among the \acp{RN} w.r.t. the model parameters received in any communication round, we analyze the probabilistic characteristics of the received model parameters to establish a fundamental understanding of how empirical risks are influenced in the face of attacks relying on shared model updates (\eg, GIA).

%

\begin{restatable}{theorem}{probfullmodel}\label{th:prob_full_model}
    For any $i,j\in[n]$ with $i\neq j$ and $t>0$, $\mathbb{P}\left[\left\lvert\theta^{(t)}_{i\not\leftarrow j}\right\rvert=0\right]$, %
    is a decreasing function of $k$.
\end{restatable}

\begin{remark}\label{rem:prob_full_model}
\Cref{th:prob_full_model} implies that under \sys, the probability of the entire model of any \ac{RN} being shared with another \ac{RN} in any communication round decreases with an increase in the number of \acp{VN}.
The proof is postponed to Appendix~\ref{sec:proof:full_model_sharing}. 
\end{remark}

\begin{restatable}{theorem}{expectedparams}\label{th:expected_params}
    For any $i,j\in[n]$ with $i\neq j$ and $t>0$, $\mathbb{E}\left[\left\lvert\theta^{(t)}_{i\not\leftarrow j}\right\rvert\right]$
    is an increasing function of $k$.
\end{restatable}

\begin{remark}\label{rem:expected_params}
\Cref{th:expected_params} ensures that under \sys, the expected number of model parameters received by any \ac{RN} from any other \ac{RN} decreases with an increase in the number of \acp{VN}.
The proof is postponed to \Cref{sec:proof:expected_params}. 
\end{remark}

In order to formalize the impact of \sys on the privacy of the shared model parameters between the \acp{RN} from an information theoretical perspective, we now analyze the \ac{MI}~\cite{ShannonInfoTheory1948} (the definition is provided in \Cref{sec:proof:mutual_info}) between the model parameters that are shared (observation of an attacker) and \emph{not} shared (secrets) between any pair of \acp{RN}. \ac{MI} and its close variants (\eg conditional entropy) have been shown to nurture a compatible relationship with formal privacy guarantees like \ac{DP}~\cite{CuffDPInforTh2016} and have been shown to capture an operational interpretation of an attacker model~\cite{KopfInforThSCA2007}. \ac{MI} measures the correlation between observations and secrets and its use as a metric to provide an information theoretical understanding of privacy is widespread in the literature. Some noteworthy examples include: gauging anonymity~\cite{ZhuInfoLeakage2005, ChatzikokolakisNoisyChannels2008}, estimating privacy in training \ac{ML} models with a cross-entropy loss function~\cite{AbadiCommWithAdvCrypt2016, TripathyPPAdvNet19, RomanelliObfMechML2020, HuangContextAwareGenAdvPrivacy2019}, and assessing location-privacy~\cite{OyaMILocationPrivacy2017, biswasprivic2023}.

\begin{assumption}\label{assump:MI_1}
    For any $i,j\in[n]$ and $t>0$, $\theta^{(t)}_j\sim\mathcal{N}\left(\vb*{\mu}^{(t)}_j,\Sigma^{(t)}_j\right)$ for some mean $\vb*{\mu}^{(t)}_j\in\mathbb{R}^d$ and covariance matrix $\Sigma^{(t)}_j\in\mathbb{R}^{d\times d}$.  
\end{assumption}

\begin{assumption}\label{assump:MI_2}
     There exists $B\in \mathbb{R}$ such that $\left(\Sigma^{(t)}_j\right)_{pp'}\leq B$ for all $p,p'\in [d]$.
\end{assumption}

\begin{assumption}\label{assump:MI_3}
    Let $d(t)$ be the number of model parameters of $N_j$ that $N_i$ receives through the interaction of their respective \acp{VN} in round $t$. 
    Setting $X_{ij}=\theta^{(t)}_{i\leftarrow j}$ and $Y_{ij}=\theta^{(t)}_{i\not\leftarrow j}$, the eigenvalues of the Schur complement $\Sigma^{(t)}_j/\Sigma^{(t)}_{Y_{ij}}$ of the block $\Sigma^{(t)}_{Y_{ij}}$ in $\Sigma^{(t)}_j=\begin{pmatrix}
\Sigma^{(t)}_{X_{ij}} & \Sigma^{(t)}_{X_{ij}Y_{ij}}\\
\Sigma^{(t)}_{Y_{ij}X_{ij}} & \Sigma^{(t)}_{Y_{ij}}
\end{pmatrix}$ are bounded below and above by $\alpha$ and $\beta$, respectively.%
\end{assumption}

\begin{assumption}\label{assump:MI_4}
$\Sigma^{(t)}_j$, $\Sigma^{(t)}_{X_{ij}}$, and $\Sigma^{(t)}_{Y_{ij}}$ are positive-definite. 
\end{assumption}

\begin{restatable}{theorem}{mutualinfo}\label{th:mutual_info}
If \Cref{assump:MI_1,assump:MI_2,assump:MI_3,assump:MI_4} hold, for any $i,j\in[n]$ with $i\neq j$ and $t>0$, we have $I\left(\theta^{(t)}_{i\leftarrow j};\theta^{(t)}_{i\not\leftarrow j}\right)\leq \Gamma \hat{B}^{d(t)} d(t)^{d(t)/2}$,

    where $\Gamma= \left(\frac{\alpha}{\beta}\right)^{\frac{\operatorname{tr}\left(\Sigma^{(t)}_j/\Sigma^{(t)}_{Y_{ij}}\right)}{\beta-\alpha}}$ and $\hat{B}=B\left(\frac{\beta^{\alpha}}{\alpha^{\beta}}\right)^{\frac{1}{\beta-\alpha}}$.
\end{restatable}

\begin{remark}\label{rem:mutual_info}
Due to \Cref{th:mutual_info}, \sys guarantees that, for a certain pair of \acp{RN} $N_i$ and $N_j$, \ac{MI} between the model parameters of $N_j$ that are shared and \emph{not} shared with $N_i$ in a certain round of communication of their respective \acp{VN} is bounded above by an increasing function of the number of parameters of $N_j$'s model that is shared with $N_i$ and bounded below by $0$ by the definition of \ac{MI}. Furthermore, recalling that \Cref{th:expected_params} ensures that the expected number of model parameters shared between any pair of \acp{RN} decreases with an increase in the number of \acp{VN}, we essentially guarantee that in every round of \sys, \ac{MI} decreases in expectation %
and, thus, the information-theoretical privacy guarantees of the \acp{RN} formally get stronger as the number of \acp{VN} increases. The proof of \Cref{th:mutual_info} is postponed to \Cref{sec:proof:mutual_info}.
\end{remark}

One of the important implications jointly posed by Theorems \ref{th:prob_full_model}, \ref{th:expected_params}, and \ref{th:mutual_info}, as highlighted by \Cref{rem:mutual_info}, is that with an increase in the number of \acp{VN}, the chances of an attacker to receive the full model of any user and the average number of model parameters exchanged between the \acp{RN} will both decrease and, consequently, the information-theoretical privacy guarantees (parameterized by \ac{MI}) for the \acp{RN} will improve. These results, in turn, provide analytical insight into how \sys is more resilient towards attacks that rely on exploiting the shared model parameters or gradient updates in their entirety. In particular, the conclusions drawn from Theorems \ref{th:prob_full_model}, \ref{th:expected_params}, \ref{th:mutual_info}, and \Cref{rem:mutual_info}, provide an information-theoretical understanding of how \sys effectively safeguards against the gradient inversion attack, as evidenced by the experimental results in \Cref{fig:exp_diff_k_images}. The findings indicate the diminished effectiveness of gradient inversion attacks with an increasing number of \acp{VN}.

\begin{remark}\label{rem:collusion}
Although the privacy of \sys is studied under the assumption of no collusion in the network, it is noteworthy that the preceding analysis can be extended to scenarios where HbC \acp{RN} collude to compromise the privacy of another \ac{RN}. Colluding \acp{RN} targeting a victim \ac{RN} $N_j$ can be modelled as $N_j$ making larger chunks of its model. As shown above, \sys enhances $N_j$'s privacy as the number of \acp{VN} increases. Thus, \sys would continue to offer better privacy guarantees compared to \ac{EL} as long as some level of chunking is maintained, with the worst-case (where all \acp{RN} collude against $N_j$) being equivalent to that of \ac{EL}. %
\end{remark}

\subsection{Overhead of \sys}
\label{app:communication_operation_cost}
Compared to standard \ac{DL} algorithms such as \dpsgd, \sys incurs some communication and operational overhead.
This can be considered as the price one has to pay for the privacy protection offered by \sys.
We now analyze these costs.

\paragraph{\textbf{Communication Costs.}}
We first compare the communication cost of \sys with that of \dpsgd, in terms of parameters sent.
In line with the rest of the paper, we use $ r $ to refer to the topology degree, $ k $ indicates the number of \acp{VN} each \ac{RN} deploys, and $ d $ indicates the number of parameters in a full model.

For simplicity, we derive the communication cost from the perspective of a single node.
In \dpsgd, a node sends $ d $ parameters to $ r $ nodes, incurring a per-node communication cost of $ dr $.
Thus, the communication cost of \dpsgd (and other standard \ac{DL} algorithms) scales linearly in the model size and topology degree.

In \sys, \acp{VN} are controlled and hosted by their corresponding \ac{RN}.
Therefore, the communication perspective of \ac{RN} should include the cost of \ac{RN}--\ac{VN} communication as well.
We separately analyze the communication cost in each of the following three phases of the \sys workflow.
\begin{enumerate}
    \item \textbf{\ac{RN} $ \rightarrow $ \acp{VN}. } In \sys, each \ac{RN} first sends a model chunk of size $ \frac{d}{k} $ to each of its $ k $ \acp{VN}, resulting in a communication cost of $ k \frac{d}{k} = d $ for this step.
    \item \textbf{\acp{VN} $ \rightarrow $ \acp{VN}. } Next, the $ k $ \acp{VN} operated by an \ac{RN} send a model chunk of size $ \frac{d}{k} $ to $ r $ other \acp{VN}, incurring a communication cost of $ k \frac{d}{k} r = dr $.
    \item \textbf{\acp{VN} $ \rightarrow $ \ac{RN}. } Finally, all $ r $ \acp{VN} who received a chunk in (2) forward these model chunks back to their associated \ac{RN}. This incurs the same communication cost as step (2):  $ dr $. %
\end{enumerate}

Therefore, the communication cost of \sys is: $ d + 2dr $.

The overhead of \sys compared to \dpsgd, in number of parameters sent, is as follows:
$$ \frac{\text{comm. cost \sys}}{\text{comm. cost \dpsgd}} = \frac{d + 2dr}{dr} = 2 + \frac{1}{r} $$

The communication cost of \sys is between $2$ and $2.5\times$ that of \dpsgd.
We do not account for the case where $r = 1$, as such a communication graph would not be connected.

\paragraph{\textbf{Number of Messages.}}
We now analyze the total number of messages sent in a single round, from the perspective of a single \Ac{RN}.
In a round of \dpsgd, a \Ac{RN} sends $ r $ messages, one to each of its neighbors.
In a round of \sys, an \Ac{RN} first sends a single message to its $ k $ \acp{VN}, \acp{VN} exchange $ kr $ messages and \acp{VN} finally forward $ kr $ received messages back to their \Ac{RN}, resulting in a total of $ k + 2kr $ messages sent.
The overhead of \sys compared to \dpsgd, in number of messages sent, is as follows:

$$ \frac{\text{num. msgs. \sys}}{\text{num. msgs. \dpsgd}} = \frac{k + 2kr}{r} = 2k + \frac{k}{r} $$

The %
\Acp{VN} in \sys introduce a level of indirection for model transfers, affecting the total communication time in each round.
While in \dpsgd models are directly exchanged between \Acp{RN}, \sys routes model chunks through \Acp{VN}, resulting in additional communication latency. %
This latency becomes more pronounced if \acp{VN} are geographically distant from their parent \acp{RN}.
The exact increase in latency depends on the underlying network infrastructure and bandwidth capabilities.
However, we remark that the time to convergence in \ac{DL} is dominated by the gradient updates and the training is not latency-critical.
Furthermore, this increase in the communication cost and number of messages is an upper bound and can be reduced through optimization techniques such as having \Acp{VN} aggregate chunks before forwarding them to their \ac{RN}.

\paragraph{\textbf{Operational Costs.}} 
There are also operational costs involved with hosting and operating \acp{VN}.
Since the \acp{VN} only need to forward messages and partial models, they can be hosted as lightweight containers that do not require heavy computational resources.
Nonetheless, they require an active network connection and sufficient memory to ensure correct operation.
\section{Evaluation}
\label{sec:evaluation}

Our evaluation answers the following questions:
\begin{enumerate*}
    \item How does the convergence and \ac{MIA} resilience of \sys evolve across training rounds, compared to our baselines (\Cref{sec:exp_convergence_mia_across_rounds})?
    \item How resilient is \sys against the privacy attacks in \DL (\Cref{sec:exp_privacy_of_nodes})?
    \item How does increasing the number of \acp{VN} affect the model convergence and attack resilience (\Cref{sec:exp_varying_k})?
    \item What is the contribution of each \sys component to privacy and convergence (\Cref{sec:ablation})?
\end{enumerate*}

\subsection{Experimental Setup}
\label{subsec:exp_setup}

\paragraph{\textbf{Implementation and compute infrastructure}}
We implement \sys using the DecentralizePy framework~\cite{dhasade2023decentralized} in approximately \num{4300} lines of Python 3.8.10, relying on PyTorch 2.1.1~\cite{paszke2019pytorch} to train and implement \ac{ML} models\footnote{Source code available at \url{https://github.com/sacs-epfl/shatter}.}.
Each node (\ac{RN} and \ac{VN}) is executed on a separate process, responsible for %
executing tasks independently of other nodes. 

All our experiments are executed on AWS infrastructure, and our compute infrastructure consists of \num{25} g5.2x large instances.
Each instance is equipped with one NVIDIA A10G Tensor Core GPU and \num{8} second-generation AMD EPYC 7R32 processors.
Each instance hosts between 2-4 \acp{RN} and 1-16 \acp{VN} for each experiment.

\paragraph{\textbf{\DL algorithm and baselines}}
\sys uses \ac{EL} as the underlying learning algorithm as \ac{EL} randomizes its communication topology every round. %
As such, we compare the privacy guarantees offered by \sys against those of \ac{EL}, \ie, in a setting without additional privacy protection.
While many privacy-preserving mechanisms are designed for \ac{FL}, only a few are compatible with \ac{DL}.
We use \muffliato as a baseline for privacy-preserving \ac{DL}~\cite{cyffersMuffliatoPeertoPeerPrivacy2022}.
\muffliato is a noise-based privacy-amplification mechanism in which nodes inject local Gaussian noise into their model updates and then conduct \num{10} gossiping rounds to average the model updates in each training round.
For \muffliato, we randomize the communication topology in each gossiping round.
While this results in $10\times$ more topology refreshes in \muffliato compared to \EL and \sys, we consider this advantage as a fair comparison.
For all experiments, unless otherwise stated, we set the number of \acp{VN} per \ac{RN} $k = $ \num{8} and generate a random regular graph with degree $r = $ \num{8} every communication round.
We assume that all nodes are online and available throughout the training.
\Cref{sec:node_dropouts} contains additional experiments with node churn, \ie, intermittent node availabilities.

\paragraph{\textbf{Privacy attacks}}
We evaluate \sys against
\begin{enumerate*}[label=\emph{(\roman*)}]
    \item Loss-based \ac{MIA},
    \item \ac{GIA}, and
    \item \ac{LA}.
\end{enumerate*}
\Ac{MIA} is a representative inference attack, quantified using the negative of the loss values on the training samples of a node and the test set of the given dataset~\cite{yeom2018privacy}.
Since \sys works with chunks of the model, \ac{LA} measures how much a \ac{VN} links to the training distribution~\cite{lebrun2022mixnn} by evaluating each received model update and evaluating the loss on the training sets of each node.
For both \ac{MIA} and \ac{LA}, one requires the full model to compute the loss.
Therefore, in \sys, the adversary complements each chunk of the received model update with the average model of the previous round to approximate the full model before attacking it.
The \ac{GIA} over the sparsified model updates is performed using the \textit{state-of-the-art} \ROG~\cite{yue2023gradient}.

\paragraph{\textbf{Learning tasks and hyperparameters}}
In our testbed, we use three datasets: \cifar~\cite{krizhevsky2014cifar}, \twitter~\cite{leaf, go2009twitter}, and \movielens~\cite{grouplens:2021:movielens} for convergence, \ac{MIA}, and \ac{LA}.
Additionally, we perform \ac{GIA} (\ROG) on \num{1600} images from the ImageNet~\cite{deng2009imagenet} validation set.

The \cifar dataset is a standard image classification task.
For \cifar, we adopt a non-IID data partitioning scheme, assigning training images to \num{100} \DL nodes with a Dirichlet distribution.
This is a common scheme to model data heterogeneity in \ac{CML}~\cite{wang2020tackling,Gong_Sharma_Karanam_Wu_Chen_Doermann_Innanje_2022,hsu2019measuring,devos2023epidemic}.
This distribution is parameterized by $ \alpha $ that controls the level of non-IIDness (higher values of $ \alpha $ result in evenly balanced data distributions).
For \cifar, we use $ \alpha = 0.1 $, which corresponds to a high non-IIDness.
This simulates a setting where the convergence is slower in \DL and the \ac{MIA} and \ac{LA} adversary have an advantage because the local distributions are dissimilar.
This task uses a \resnet model (with $\approx$\num{11.5} million parameters)~\cite{he2016deep}.

Our second dataset is \twitter, a sentiment classification task containing tweets partitioned by users~\cite{go2009twitter, leaf}.
We preprocess the dataset such that each user has at least \num{24} tweets.
All tweets have been annotated with a score between 0 (negative) and 4 (positive).
We randomly assign the different authors behind the tweets to \num{50} \ac{DL} nodes, a partitioning scheme that is also used in related work on \ac{DL}~\cite{dhasade2023get,dhasade2023decentralized}.
For this task, we initialize a pre-trained \textsc{bert-base-uncased} Transformer model (with $ \approx $\num{110} million parameters)~\cite{devlin2018bert} from Hugging Face with a new final layer with \num{2} outputs.
We then fine-tune the full model on the \twitter dataset.

We also evaluate \sys using a recommendation model based on matrix factorization~\cite{korenmatrixfactorization2009} on the \movielens 100K dataset, comprising user ratings from \num{0.5} to \num{5} for several movies.
This task models a scenario where individual participants may wish to learn from the movie preferences of other users with similar interests.
Like the \twitter task, we consider the multiple-user-one-node setup where we randomly assign multiple users to a \ac{DL} node.

Lastly, to present the vulnerability of \DL and \sys against \ac{GIA}, we instantiate \num{100} \DL nodes, each with \num{16} images from the ImageNet validation set.
We use the \lenet model to perform one \DL training round.
\ROG is performed on the model updates exchanged after the first training round.
Since the reconstructed images are the best after the first training round, we do not continue further \DL training and evaluate the strongest version of this attack.

We perform \num{1} epoch of training before exchanging models for all convergence tasks and \num{5} epochs of training for the ImageNet \ac{GIA}.
Learning rates were tuned on a grid of varying values.
For \muffliato, we comprehensively use two noise levels: 
\begin{enumerate*}[label=\emph{(\roman*)}]
    \item low: with good convergence, which results in low privacy against \ac{MIA}, and %
    \item high: with higher attack resilience to \ac{MIA}, but at the cost of model utility. %
\end{enumerate*}
Lower noise values present no privacy benefits against \EL, and higher noise values adversely affect convergence.
More details about the used datasets and hyperparameters can be found in \Cref{app:exp_setup}.
In summary, our experimental setup covers datasets with differing tasks, data distributions, and model sizes, and the hyperparameters were carefully tuned.

\paragraph{\textbf{Metrics and reproducibility}}
All reported test accuracies are top-1 accuracies obtained when evaluating the model on the respective test datasets.
For \movielens, we only present the RMSE-loss, as it is not a classification task.
To quantify the attack resilience of \sys and its baselines, we use the ROC-AUC metric when experimenting with the \ac{MIA}, representing the area-under-curve when plotting the true positive rate against the false positive rate.
For the \ac{LA}, we report the success rate in percentage, \ie, for each adversary, the success rate is %
the percentage of \textit{correctly linked} model chunks out of the \textit{total attacked} model chunks.
As the attacks are extremely computationally expensive, each \ac{RN} across all baselines randomly evaluates \num{8} received model updates for \cifar and \movielens, and \num{4} for the \twitter dataset every training round.
Finally, we report the LPIPS score when evaluating \ROG, a score between 0 and 1, which indicates the similarity between two batches of images (more similar images correspond to lower LPIPS scores).
We run all experiments three times, using different seeds, and present averaged results with confidence intervals.

\subsection{Convergence and \Ac{MIA} Across Rounds}
\label{sec:exp_convergence_mia_across_rounds}
We first evaluate how the convergence and resilience against the \ac{MIA} evolve with the training for \sys and the baselines for all three datasets.
Each \ac{RN} operates \num{8} \acp{VN}, and we train until \EL convergence.
An ideal algorithm would have a low \ac{MIA} AUC and a high test accuracy (or low test loss for \movielens) close to \EL with no privacy-preserving solution.

\begin{figure}[t!]
	\centering
	\inputplot{plots/across_rounds_mlens}{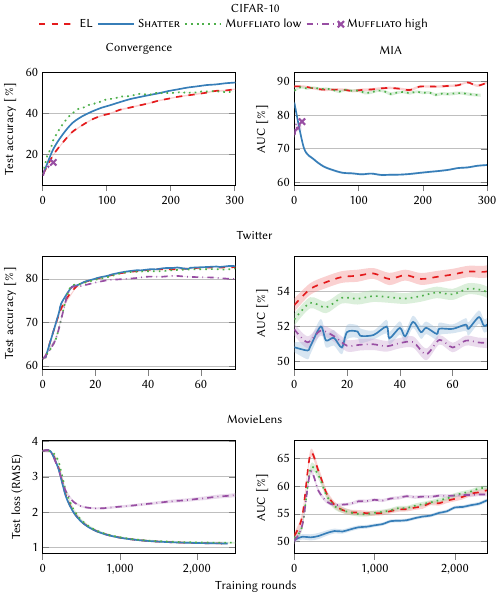}
	\caption{The test accuracy ($\uparrow$ is better) and \ac{MIA} AUC ($\downarrow$ is better) across training rounds for \cifar (top), \twitter (middle) and the test loss ($\downarrow$ is better) for \movielens (bottom), for \ac{EL}, \sys and \muffliato. An AUC of 50\% corresponds to an attacker randomly guessing. } %
	\label{fig:across_rounds_mlens}
\end{figure}

\Cref{fig:across_rounds_mlens} shows the convergence (left) and \ac{MIA} attack success (right) for each dataset (row-wise).
\EL achieves \num{51.74}\%, \num{83.11}\%, and \num{1.10}~[RMSE] model utility for \cifar, \twitter, and \movielens, respectively.
On the other hand, \EL is vulnerable to the loss-based \ac{MIA}, reaching close to \num{89.7}\% AUC on \cifar, \num{56.1}\% AUC on Twitter, and \num{65.8}\% AUC on \movielens respectively, where the chance of a random guess is \num{50}\%.
\muffliato~(\textit{low}) achieves similar convergence to \EL for all datasets but is also not very effective at reducing vulnerability against the \ac{MIA}.
\muffliato~(\textit{high}) instead lowers the \ac{MIA} vulnerability but at the cost of convergence, finally reaching \num{2.61}\% lower test accuracy for \twitter, and \num{1.0}~[RMSE] higher test loss for \movielens (considering the best values).
For the \cifar dataset, convergence is so unstable that there is no learning beyond \num{12} training rounds.
In contrast, \sys consistently outperforms the baselines in terms of the resilience to \ac{MIA} while having no adverse impact on convergence for \twitter and \movielens, and positively raises the top-1 test accuracy by \num{3.21}\% for \cifar.
For a given iteration, \sys reduces \ac{MIA} AUC~(\%) by \num{4.9}-\num{25.8}\% on \cifar, \num{3.4}-\num{4.0}\% on \twitter, and \num{0.8}-\num{15.1}\% on \movielens.

These experiments cover a range of settings, \eg, we find that the extreme \niid{}ness in \cifar adversely affects \muffliato, whereas it positively affects \sys.
Furthermore, the pre-training in \twitter experiments makes it more challenging for the adversary to perform a \ac{MIA} as models become less personalized.
In \cifar, \sys has a higher vulnerability to \ac{MIA} at the start, decreasing sharply afterward.
We notice the accuracy of the models at this point is as good as a random guess (\num{10}\%), and hence training techniques like regularization can fix it.
However, we do not evaluate with regularization and show the worst case from the attack's perspective.
Another seemingly unexpected behavior is the higher vulnerability of \muffliato~(\textit{high}) against \muffliato~(\textit{low}) for the \movielens dataset.
This can be explained by the loss curve on the left, which clearly shows significantly worse generalization for \muffliato~(\textit{high}) than \muffliato~(\textit{low}).

\subsection{Privacy Guarantees of Individual Nodes}
\label{sec:exp_privacy_of_nodes}

\begin{figure}[t!]
	\centering
	\inputplot{plots/across_clients_mlens}{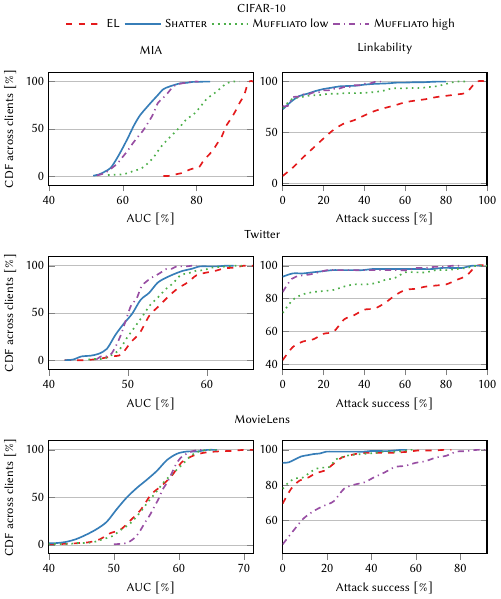}
	\caption{The distribution of \ac{MIA} AUC ($\downarrow$ is better) and attack success of the \ac{LA} across clients for \cifar (top), \twitter (middle) and \movielens (bottom)
 , for \ac{EL}, \sys and \muffliato. 
 An attack success of 2\% for the \ac{LA} corresponds to an attacker randomly guessing on \twitter and 1\% on other datasets.}%
	\label{fig:clients_cdfs_mlens}
\end{figure}

We now analyze the vulnerability of individual nodes to the \ac{MIA} and \ac{LA}.
Ideally, we aim for most nodes to have a low \ac{MIA} AUC and \ac{LA} success.
\Cref{fig:clients_cdfs_mlens} shows the CDF of victim nodes against the average \ac{MIA} AUC (left) and \ac{LA} success (right) for the baselines.
We observe that, consistently across datasets, \sys outperforms \muffliato~(\textit{low}) and \EL in defending against both attacks.
\muffliato~(\textit{high}) appears to beat \sys on the \twitter dataset and appears to be a better defense against \ac{LA} for \cifar.
However, we need to keep in mind that \muffliato~(\textit{high}) significantly hurts convergence (see \Cref{fig:across_rounds_mlens}).

\Ac{LA} %
also assesses the vulnerability of the real identities of the nodes that can be leaked through their local data distributions, even when the real identities and model updates are obfuscated.
\sys preserves the privacy of the nodes by having near-zero linkability for at least \num{70}\% across our experiments, up from only \num{7}\% nodes with near-zero linkability in \EL.
Hence, \Cref{fig:clients_cdfs_mlens} (right) demonstrates that just obfuscating identities, \eg, by using onion routing techniques~\cite{goldschlag1999onion}, would not defend against \ac{LA}.
While \muffliato~(\textit{high}) is also decent in the defense against \ac{LA}, this needs to be looked at in conjunction with the convergence.
To conclude, \sys offers individual clients better privacy protection compared to the baselines and across a range of varying learning tasks.

\begin{figure*}[ht!]
	\centering
	\inputplot{plots/box_plots}{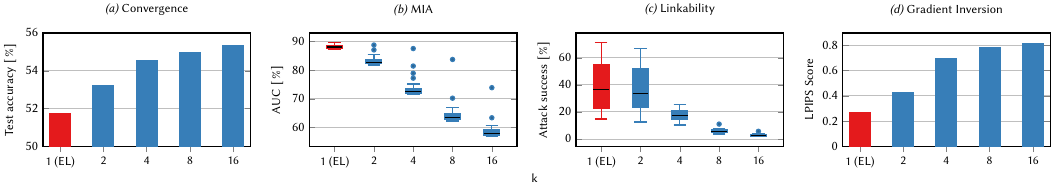}
	\caption{The test accuracy (a, $\uparrow$ is better), \ac{MIA} AUC (b, $\downarrow$ is better), attack success rate for \ac{LA} (c, $\downarrow$ is better) on \cifar, and \ac{GIA} LPIPS score (d, $\uparrow$ is better) on ImageNet for increasing values of $ k $ (with $ k = 1 $ corresponding to an \ac{EL} setting).}
    \label{fig:exp_diff_k}
\end{figure*}

\subsection{Varying the Number of \acp{VN}}
\label{sec:exp_varying_k}
We next analyze the impact of an increased number of \acp{VN} on convergence and resilience against our three privacy attacks.
As we increase the number of \acp{VN} per \ac{RN} ($k$), we keep the degree of the topology constant ($r = 6$), consequently keeping the total bytes transferred constant for \sys.
We establish the communication and operational costs of \sys in \Cref{app:communication_operation_cost}.

\paragraph{\textbf{Convergence}}
To observe the impact of $ k $ on convergence, we run \ac{EL} and \sys with increasing values of $ k $ for \num{300} training rounds with the \cifar dataset.
The convergence bar plot in~\Cref{fig:exp_diff_k}a shows the average test accuracy during a run for different values of $ k $.
\ac{EL} achieves an average test accuracy of 51.7\%, which increases to 53.2\% with $ k = 2 $.
Further increasing $ k $ positively affects the test accuracy: increasing $ k $ from \num{2} to \num{16} increases the attained test accuracy from 53.2\% to 55.3\%.
We attribute this increase in test accuracy to the superior propagation and mixing of model chunks.
Furthermore, this increase in test accuracy and faster convergence compensates for the increased communication cost of \sys.

\begin{figure}[b!]
	\centering
 \includegraphics[width=\linewidth]{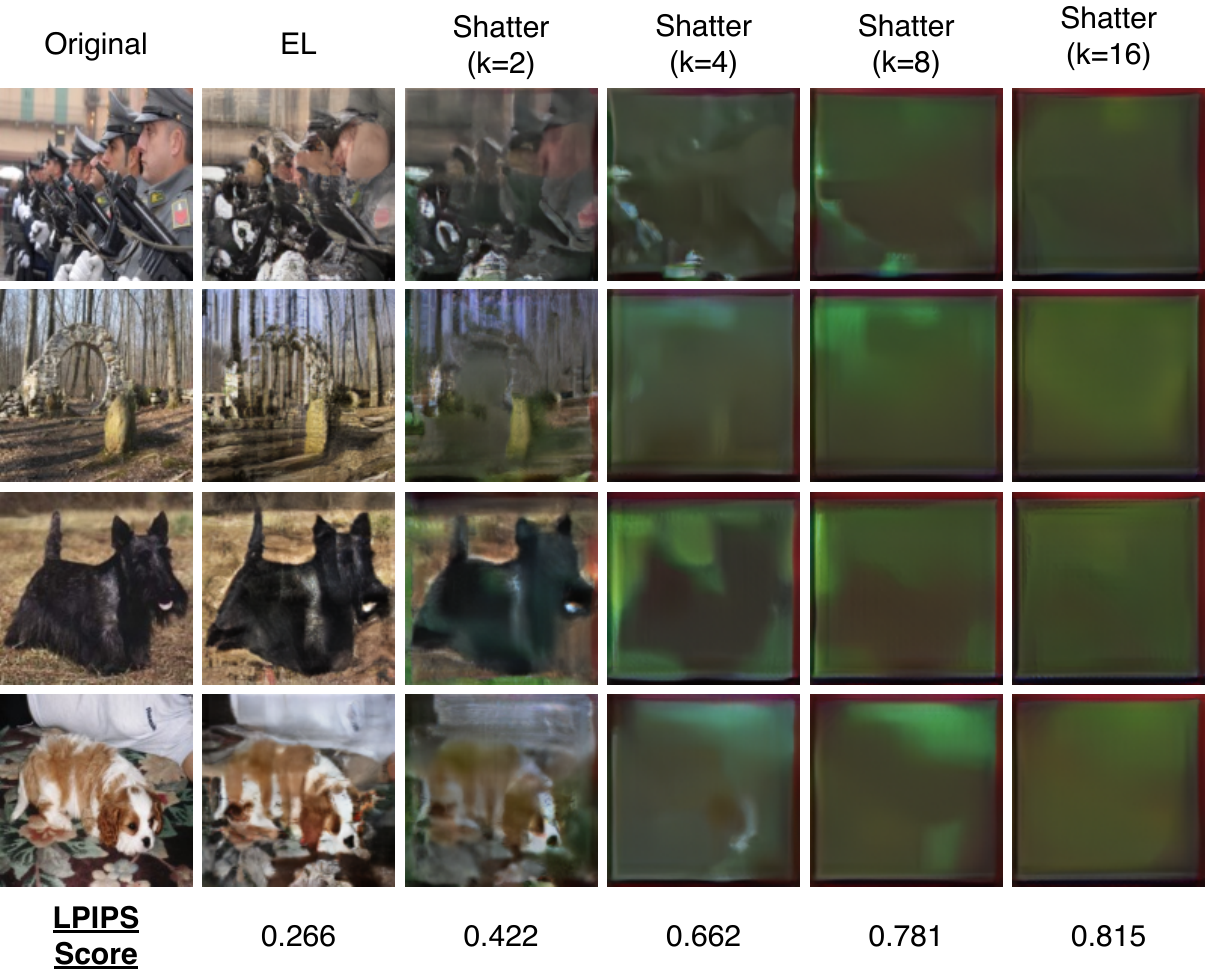}
	\caption{Selected reconstructed images using the \ac{GIA}, for our baselines and different numbers of \acp{VN} ($k$). We also show the average LPIPS scores ($\uparrow$ is better) for all 1600 reconstructed images for each setting at the bottom of the figure.}
	\label{fig:exp_diff_k_images}
\end{figure}

\paragraph{\textbf{Attack resilience}}
Next, we compare the attack resilience of \sys with different values of $ k $ against that of \ac{EL}, for all three privacy attacks.
Figures \ref{fig:exp_diff_k}b-d show the LPIPS score, AUC, and attack success percentage for the \ac{MIA}, \ac{LA}, and \ac{GIA} respectively, for increasing values of $ k $. %
The values for a fixed $ k $ are averaged across clients and training rounds.
\Cref{fig:exp_diff_k}b shows that %
$ k = 2 $ already defends better against \ac{MIA} and reduces the median AUC from 88\% to 83\%.
\sys with $ k = 16 $ provides significant privacy guarantees, showing a median AUC of only 58\%.
A similar trend is visible for the \ac{LA}, as shown in~\Cref{fig:exp_diff_k}c, where an attacker has a success rate of 36.5\% in \ac{EL} and it decreases rapidly with an increase in $k$.
For $ k = 16 $, a median attack success of merely 2.5\% (and at most 4.5\%) is observed, underlining the superior privacy benefits of \sys.

We conduct the \ac{GIA} on ImageNet when the network learns using a \lenet model, using the experimental setup described in~\Cref{subsec:exp_setup}.
Four random images from a set of $1600$ reconstructed images are shown in~\Cref{fig:exp_diff_k_images}, with the original image in the left-most column and the images for \ac{EL} and \sys for different values of $ k $ in the other columns.
For each approach, we also show the average LPIPS score~\cite{zhang2018unreasonable} over all the images in \Cref{fig:exp_diff_k}d.
For \ac{EL}, we observe significant similarities between the original and reconstructed images with a low LPIPS score.
For $ k \geq 4 $, the reconstructed images increasingly become too blurry to visually obtain any meaningful information, as evidenced by an increase in the LPIPS score.
For $ k = 16 $, we cannot identify any semantic features in the reconstructed image with a $3.1\times$ higher LPIPS score compared to \EL.
Employing more \acp{VN}, \eg, increasing $ k $, thus results in additional privacy.
We discuss the choice of $k$ in more detail in \Cref{sec:final_remarks}.

\begin{figure}[tb]
	\centering
	\inputplot{plots/ablation2.tex}{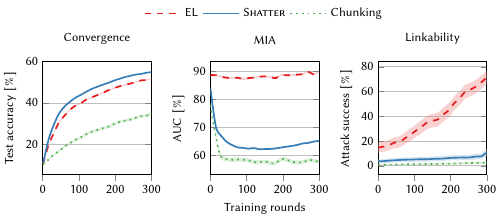}
	\caption{Ablation study: convergence and resistance against \ac{MIA} and \ac{LA} %
        with different components of \sys enabled.}
	\label{fig:ablation}
\end{figure}

\subsection{Ablation Study}
\label{sec:ablation}
We now evaluate the effectiveness of the components of \sys, \ie, \emph{chunking}, \emph{full sharing} and \emph{virtualization}, in terms of convergence and resistance against the \ac{MIA} and \ac{LA}.
\Cref{fig:ablation} shows the performance of \textit{chunking}, \ie, sharing one chunk with $1/8^{\operatorname{th}}$ of model parameters and \sys (\textit{full sharing} + \textit{chunking}) against that of \EL.
We omit \textit{virtualization} (unlinking identities) in this experiment as it %
does not affect convergence.
Furthermore, the privacy benefits of \textit{chunking} diminish without \textit{virtualization} because an attacker would trivially be able to put together two chunks received from the same source. %
We observe that \textit{chunking} improves privacy w.r.t. both \ac{MIA} and \ac{LA} at the cost of accuracy.  %
With only \textit{chunking}, the nodes reach \SI{16.9}{\%} lower accuracy points than \EL and \SI{19.8}{\%} lower accuracy points than \sys with 8 \acp{VN}. 
When comparing \sys and \textit{chunking} for privacy, the lower attack success for \textit{chunking} is due to the restrictive learning rate.
We tuned the learning rate for optimal convergence and the optimal learning rate for \textit{chunking} is much lower than that of \sys.
This results in smaller model updates with \textit{chunking} as opposed to \sys, making \ac{MIA} and \ac{LA} more difficult to carry out.
In summary, we elucidate the contributions of the components of \sys: \textit{chunking} and \textit{virtualization} bring privacy, and \textit{full sharing} brings utility without compromising the efficiency compared to \EL.
\section{Related Work}
\label{sec:prelims_solutions}

\paragraph{\textbf{Trusted hardware}}
Trusted execution environments (TEE) create secure environments on the processors to safeguard data and calculations from untrusted administrative domains~\cite{sabt2015trusted}.
Systems such as ShuffleFL~\cite{zhang2021shufflefl}, Flatee~\cite{mondal2021flatee}, and Papaya~\cite{huba2022papaya} use secure hardware for private averaging to prevent the server from inspecting model updates.
GradSec~\cite{messaoud2022shielding} uses ARM TrustZone to prevent inference attacks in \ac{FL}.
ReX~\cite{dhasade2022tee} uses Intel SGX for secure data sharing in decentralized learning.
While these systems effectively hide model updates from the server operator or nodes, they require specialized hardware that is not always available.
\sys, however, enhances privacy without needing any specialized hardware.

\paragraph{\textbf{Secure aggregation}}
Secure aggregation is a privacy-enhancing method to securely share models using masks and has been deployed in the context of \ac{FL}~\cite{bonawitz2017practical,fereidooni2021safelearn}.
These masks are agreed upon before training and cancel out upon aggregation.
This scheme requires interactivity and sometimes extensive coordination among participants, requiring all nodes to apply masks, be connected and available. %
\sys avoids using cryptographic techniques and has less strict requirements on coordination among nodes.

\paragraph{\textbf{Differential privacy}}
Several differential privacy schemes obtain provable privacy guarantees in \ac{DL}~\cite{cyffersMuffliatoPeertoPeerPrivacy2022, lin2022towards}.
These works add %
noise %
to model weights before sharing them with the neighbors.
While they provide strong theoretical privacy guarantees, we have demonstrated that they often significantly deteriorate model utility.
Nevertheless, privacy-critical applications benefit from differential privacy guarantees.
In contrast, \sys does not introduce such noise and, therefore, avoids the negative effect on model utility.

\paragraph{\textbf{Onion routing and \acp{VN}}}
The notion of \acp{VN} in 
\sys somewhat resembles onion routing, which hides a sender's identity by routing traffic through intermediate nodes~\cite{goldschlag1999onion}.
While onion routing techniques can be applied to \ac{DL}, it would not defend %
against \ac{MIA} and \ac{GIA}, and would be innocuous against \ac{LA} %
(\cf \Cref{fig:exp_diff_k}). %
\acp{VN} were also used in P2P networks, \eg, to ensure load balancing in DHTs~\cite{awad2008virtual}.
On the other hand, \acp{VN} in \sys exchange models on behalf of \acp{RN} improving the privacy and accuracy of \ac{DL}.

\section{Final Remarks}
\label{sec:final_remarks}
\sys is a novel %
approach to privacy-preserving \ac{DL}.
It %
partitions models into chunks and distributes %
them across a dynamic communication topology of virtual nodes that significantly enhances privacy by preventing adversaries from reconstructing complete models and from identifying the original nodes responsible for specific %
contributions.
We have theoretically and empirically demonstrated the convergence of \sys, its formal privacy guarantees, and its resilience against three state-of-the-art privacy attacks across a variety of learning tasks.

\paragraph{\textbf{Choosing \textit{k}}}
The number of \acp{VN} per \ac{RN}, $ k $, is a free parameter in \sys.
The appropriate value of $ k $ in a deployment setting depends on the specifications of the available infrastructures and the sensitivity of the training data %
as the choice of $ k $ directly impacts privacy guarantees and overhead.
While environments with powerful infrastructures can use higher values of $k$ to enhance privacy, %
a lower $k$ may be more appropriate in managing the overhead for settings with limited computational resources. Additionally, highly sensitive data benefits from a higher $k$, even if this results in more overhead.
The choice of $ k $ being highly domain- and application-specific, our experiments show that $ 8 \leq k \leq 16 $ %
strikes a reasonable balance between privacy guarantees and overhead.

\paragraph{\textbf{\sys and differential privacy}}
\sys protects the privacy of individual model updates shared in \DL.
However, the final model may still leak some information.
While this threat model is widely adopted in related work~\cite{huba2022papaya,bonawitz2017practical,fereidooni2021safelearn},
to obtain DP guarantees for extra privacy on the final model, one may inject noise to model parameters before \emph{chunking}. 
This is, however, orthogonal to \sys and comes at the cost of utility.

\paragraph{\textbf{\sys and sparsification}}
The communication between two \acp{VN} resembles random sparsification.
\sys leverages the fact that it is more difficult to attack a partial set of parameters than the full model.
From the perspective of a single \ac{RN}, \sys still disseminates as much information as \DL (see \Cref{fig:shatter_is_not}, \Cref{app:dpsgd} for details).
We remark that \sys is compatible with sparsification for improving communication efficiency.
For instance, the \ac{RN} may still use the state-of-the-art sparsification schemes on the trained model parameters and perform \textit{chunking} only on the sparsified parameters. We believe such an analysis would be interesting as a potential future avenue to generalize \sys. 

\paragraph{\textbf{Active attacks}}

While this work focuses on a HbC threat model with a passive adversary, some networks may contain active adversaries that deviate from the specified protocol, \eg, by modifying model parameters or colluding with other nodes.
A general defense mechanism against active attacks is difficult to realize due to the large design space of such attacks.

Nevertheless, the components of \sys raise the bar to mount particular active attacks.
Firstly, \sys randomizes the communication topology in each round, which makes it difficult for an attacker to consistently target a specific set of nodes.
The latter is a requirement for the successful execution of many active attacks~\cite{pasquini2023security}.
Secondly, the chunking process involves dividing the model into smaller parts (chunks) and distributing them among \acp{VN}.
Model chunking limits the exposure of any single part of the model, thereby reducing the potential impact of an active attack.
Thanks to the randomization in \sys, it is not guaranteed that the chunks by an active attacker end up in the model of a particular victim.
The combined effect of these components is that an attacker’s ability to mount privacy attacks is significantly crippled.
The security of \sys can be further extended with auxiliary defense mechanisms to protect against advanced active and Sybil attacks~\cite{cyffersMuffliatoPeertoPeerPrivacy2022,yu2023ironforge,gupta2023byzantine}.
We aim to explore this in future work.

\begin{acks}
This work has been funded by the Swiss National Science Foundation, under the project ``FRIDAY: Frugal, Privacy-Aware and Practical Decentralized Learning'', SNSF proposal No. 10.001.796.
Mathieu Even and Laurent Massoulié were supported by the French government under management of the ``Agence Nationale de la Recherche'' as part of the ``Investissements d'avenir'' program, reference ANR19-P3IA-0001 (PRAIRIE 3IA Institute).
We extend our gratitude to Amazon Web Services (AWS) and the ``Gesellschaft für wissenschaftliche Datenverarbeitung mbH Göttingen (GWDG)'', Germany, for providing the compute credits used to run our experiments.
\end{acks}

\bibliographystyle{ACM-Reference-Format}
\bibliography{main.bib}

\newpage

\appendix

\section{Table of Notations}\label{app:notations}

\begin{table}[h]
\centering
\begin{tabular}{|c|l|}
\hline
\textbf{Notation} & \textbf{Description} \\ \hline \hline
\multicolumn{2}{c}{D-PSGD}\\ \hline \hline
$\mathcal{N}$ & Set of all nodes\\ \hline
$\cG$ & Graph topology \\ \hline
$\cE$ & Edges (communication) between the nodes \\ \hline
$D_i$ & Local dataset of node $i$\\ \hline
$\xi_i$ & Mini-batch sampled from $D_i$ \\ \hline
$\theta^{(t)}_i$ & Local model of node $i$ in round $t$\\ \hline
$f_i$ & Loss function of node $i$\\\hline
$H$ & Number of local training steps \\ \hline \hline
\multicolumn{2}{c}{\sys}\\ \hline \hline
$\mathcal{N}$ & Set of all \acp{RN} \\ \hline
$n$ & Number of \acp{RN} (= $|\mathcal{N}|$) \\ \hline
$N_i$ & $i^{\operatorname{th}}$ \ac{RN} \\ \hline
$k$ & Number of \acp{VN} per \ac{RN} \\ \hline
$r$ & Number of neighbors of each \ac{VN} \\ \hline
$d$ & Dimension of the parameter space of the models \\ \hline 
$\theta^{(t)}_i$ & model held by $N_i$ in round $t$\\ \hline
$v_i(s)$ & $s^{\operatorname{th}}$ \ac{VN} of $N_i$ for all $i\in[n]$ and $s\in[k]$\\ \hline
$\theta_{i,s}^{(t)}$ & chunk of $N_i$'s model held by $v_i(s)$ in round $t$ \\ \hline 
$\theta^{(t)}_{i\leftarrow j}$ & Part of $N_j$'s model received by $N_i$ in round $t$\\ \hline
$\theta^{(t)}_{i\not\leftarrow j}$ & Part of $N_j$'s model  \emph{not} received by $N_i$ in round $t$\\ \hline
$\cG_t$ & Communication graph of the \acp{VN} in round $t$\\ \hline
$\cV$ & Set of all \acp{VN} \\ \hline
$\cE_t$ & Edges (communication) between the \acp{VN} in round $t$\\\hline
\end{tabular} 
\end{table}

\begin{table*}[t!]
\small
    \centering
    \caption{Summary of datasets used for convergence evaluation. For each dataset, the tuned values of learning rate (\(\eta\)), batch size ($\mathcal{B}$), the number of training and testing samples, total model parameters and noise-levels for \muffliato ($\sigma$) are presented.}
    \label{tab:datasets}
    \begin{tabular}{l l l r c r r r c c c}
        \toprule
        \multirow{2}{*}{\textbf{Task}} & \multirow{2}{*}{\textbf{Dataset}} & \multirow{2}{*}{\textbf{Model}} & \multirow{2}{*}{\textbf{\(\eta\)}} & \multirow{2}{*}{\textbf{$\mathcal{B}$}} & \textbf{Training} & \textbf{Testing} & \textbf{Total} & \multicolumn{3}{c}{\textbf{\muffliato Noise ($\sigma$)}} \\
         &  &  &  &  & \textbf{Samples} & \textbf{Samples} & \textbf{Parameters} & \textbf{Low} & \textbf{Medium} & \textbf{High} \\
        \midrule
        Image Classification & \cifar & \resnet & 0.050 & 32 & \num{50000} & \num{10000} & \num{11494592} & 0.025 & 0.050 & 0.100 \\ 
        Sentiment Analysis & \twitter & BERT & 0.005 & 32 & \num{32299} & \num{8484} & \num{109483778} & 0.009 & - & 0.015 \\ 
        Recommendation & \movielens & Matrix Factorization & 0.075 & 32 & \num{70000} & \num{30000} & \num{206680} & 0.025 & - & 0.250 \\ 
        \bottomrule
    \end{tabular}
\end{table*}

\begin{figure*}
	\centering
	\includegraphics{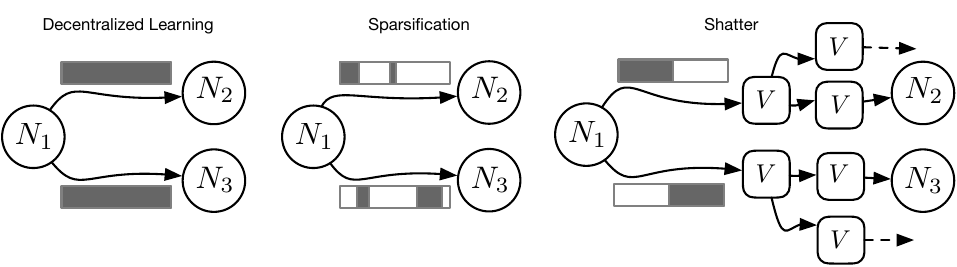}
	\caption{Distinguishing \sys from \DL and Sparsification.} %
	\label{fig:shatter_is_not}
\end{figure*}

\section{The \dpsgd Algorithm}
\label{app:dpsgd}
We show in~\Cref{alg:dpsgd} the standard \dpsgd procedure, from the perspective of node $ i $.
\Cref{fig:shatter_is_not} illustrates the differences between \dpsgd, Sparsification and \sys.

\begin{algorithm2e}[h]
    \DontPrintSemicolon
    \caption{The \dpsgd procedure, from the perspective of node $i$.}
    \label{alg:dpsgd}
    Initialize $\theta_i^{(0)}$\;
    \For{$t = 0, \dots, T-1$}{
    $\tilde \theta_i^{(t,0)}\leftarrow\theta_i^{(t)}$\;
        \For{$h = 0, \dots, H-1$}{\label{line:local_update_start}
            $\xi_i \leftarrow$ mini-batch sampled from $D_i$\;
            $\tilde\theta_i^{(t,h+1)} \leftarrow \tilde\theta_i^{(t,h)} - \eta \nabla f_i(\tilde\theta_i^{(t,h)}, \xi_i)$\;
        }\label{line:local_update_end}
        \textbf{Send} $\tilde\theta_i^{(t,H)}$ to the neighbors in topology $ \cG $ \label{line:share}\;
       \textbf{Receive} $\tilde\theta_j^{(t,H)}$ from each neighbor $j$ in $ \cG $ \label{line:receive}\;
         \textbf{Aggregate} the received models to produce $\theta_i^{(t+1)}$\label{line:aggregate}\;
    }
    \KwRet $\theta_i^{(T)}$\;
\end{algorithm2e}

\section{Alternative Model Chunking Strategy}
\label{app:alternative_chunking_strategies}
\acp{RN} in \sys break down their local model into smaller chunks that are propagated by their \acp{VN}.
The strategy for model chunking can affect both the privacy guarantees and the convergence of the training process and should therefore be carefully chosen.
As discussed in~\Cref{sec:vnodes}, each \ac{RN} randomly samples weights from its local model, without replacement.
Furthermore, all \acp{RN} create chunks with the same indices.
We acknowledge that other chunking strategies are possible; we discuss a few alternative model chunking strategies and analyze their trade-offs in terms of privacy and convergence.

\paragraph{\textbf{1) Linear chunking}} With linear chunking, the model parameters are flattened to a one-dimensional array of weights and linearly partitioned into $ k $ chunks.
Thus, the first $ \frac{1}{k} $ parameters are given to the first \ac{VN}, the next $ \frac{1}{k} $ parameters to the second \ac{VN} etc.
However, we experimentally found this strategy to violate privacy since different model chunks carry different amounts of information, \ie, they contain information from different layers.
For example, the final chunk, containing the weights associated with the final layer, can leak substantial gradient information~\cite{zhao2020idlg, wainakh2021label}.

\paragraph{\textbf{2) Random weight sampling (with replacement)}} Another strategy involves a \ac{RN} sampling $ \frac{1}{k} $ random weights, with replacement.
With this strategy, a \ac{RN} might send the same weight to multiple of its \acp{VN} during a round.
However, this risks an adversarial \ac{RN} inferring that two model chunks originate from the same \ac{RN}.
Specifically, a \ac{RN} receiving two model chunks that share a common weight at the same position can now, with a high likelihood, conclude that those chunks originate from the local model of the same \ac{RN}. %
Furthermore, this strategy might result in a situation where important weights are not sampled at all during a particular training round, which negatively affects performance.

\paragraph{\textbf{3) Dynamic chunking.}}
While we stick to \textit{static} chunking in \sys, another approach would be to change the parameter assignments for each \ac{VN} over time.
While combining this with a random chunking strategy increases the stochasticity of the system, this can potentially be worse for privacy closer to convergence.
The parameters do not change much closer to convergence. Therefore, an adversary may continue to listen to the same \ac{VN} across rounds sending different sets of parameters and may potentially collect all the model parameters of \ac{RN}.
Combining this with the fact that \ac{GIA} is highly effective close to convergence~\cite{pasquini2023security}, we may end up with the privacy vulnerability of \ac{EL}.

\section{Experimental Setup Details}
\label{app:exp_setup}
\Cref{tab:datasets} provides a summary of the datasets along with the hyperparameters used in our testbed.
Sentiment analysis over \twitter is a fine-tuning task, whereas we train the models from scratch in the other tasks.
In addition to these datasets, we use ImageNet validation set for the \ac{GIA} using \ROG.
Two nodes are assigned \num{16} images each and they train with the batch size $b = 16$ for \num{5} epochs using the LeNet model architecture.
We perturb each node's initial model randomly for increased stochasticity.

\section{Experiments with Node Dropouts}
\label{sec:node_dropouts}

We present the convergence of \sys in the setting where \DL nodes have intermittent connectivity.
We simulate different dropout rates for \DL nodes and \acp{RN} in \sys.
In each round, nodes decide to participate or not based on the dropout rate.
In addition to the dropout rate, we set a dropout correlation factor of $10\%$, \ie, a node is $10\%$ more likely to drop out if it did not participate in the last round.
When a node does not participate in a round, it does not perform training, sharing, and aggregation.
When a \ac{RN} in \sys drops out, we assume that its \acp{VN} also drop out.
While we assumed an $r-regular$ topology in our system model~(\Cref{sec:prelims_threat}), this does not hold when nodes have intermittent availabilities.
We assume that the communication topology refreshes are agnostic of the availabilities of the nodes.

\begin{figure}[t!]
	\centering
	\inputplot{plots/dropouts}{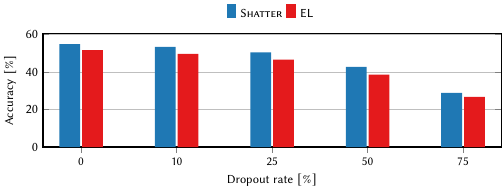}
	\caption{Accuracy achieved after 300 rounds at different dropout rates.}
\label{fig:dropouts}
\end{figure}

\Cref{fig:dropouts} shows the highest test accuracy reached after $300$ rounds of \cifar training for varying dropout rates.
In \sys, we assume $ k = 8$, \ie, $8$ \acp{VN} per \ac{RN}.
\sys consistently performs better than \ac{EL} even with the dropout rate of as high as $75\%$.
\sys with a $10\%$ dropout rate performs even better than \ac{EL} without any node dropouts.
We notice a drop in the accuracy as the drop rate goes up for both \sys and \ac{EL}.
This is expected as fewer gradients are computed and fewer parameters are aggregated because some nodes are unavailable for training and aggregation in each round.
These results demonstrate that \sys is not hampered more by node dropouts when compared to the \textit{state-of-the-art} \ac{DL} baseline of \ac{EL}.
Furthermore, the achieved model utility in \sys is not significantly affected for topologies slightly deviating from $r-regular$ as seen from the results with the dropout rate of $10\%$.

\section{Postponed proofs}

\subsection{Proof of \Cref{thm:convergence}}\label{sec:proof:convergence}

We start by proving the following Lemma. 
\begin{lemma}\label{lem:contraction}
    Let $\cC_t:\R^{d\times n}\to \R^{d\times n}$ be the operator defined as 
    \begin{equation*}
        \forall \theta=(\theta_i)_{i\in[n]}\in\R^{d\times n}\,,\quad \cC_t(\theta)_i= \sum_{s=1}^k \frac{1}{r} \sum_{w\in\cV:\{v_i(s),w\}\in\cE_t} \theta_w\,,
    \end{equation*}
    where for $w=v_i(s)$ a virtual node of $N_i$, $\theta_w\in\R^d$ contains the subset of coordinates of $\theta_i$ associated to the model chunks of $v_i(s)$.
    We have, for all $\theta=(\theta_i)_{i\in[n]}\in\R^{d\times n}$:
    \begin{align}
        \frac{1}{n}\sum_{i=1}^n\cC_t(\theta)_i&=\bar\theta\,,\\
        \mathbb{E}\left[\frac{1}{n}\sum_{i=1}^n\left\| \cC_t(\theta)_i - \bar\theta\right\|^2 \right]&\leq  \frac{\rho}{n}\sum_{i=1}^n\left\|\theta_i - \bar\theta\right\|^2\,,\label{ineq:contraction}
    \end{align}
    where $\bar\theta=\frac{1}{n}\sum_{i=1}^n\theta_i$ and
    \begin{equation}\label{eq:rho}
        \rho = \frac{nk}{r(nk-1)}\left( 1+ \frac{k-r}{nk-3} + \frac{(k-1)(r-1)}{(nk-2)(nk-3)}\right)\,.
\end{equation}
\end{lemma}

\begin{proof}[Proof of \Cref{lem:contraction}]\let\qed\relax
    Let $\theta\in\R^{d\times n}$ and $\ell\in[d]$.
    For the first part of the lemma, we have:
    \begin{align*}
        \frac{1}{n}\sum_{i=1}^n\cC_t(\theta)_i &= \frac{1}{n} \sum_{i=1}^n\sum_{s=1}^k \frac{1}{r} \sum_{w\in\cV:\{v_i(s),w\}\in\cE_t} \theta_w\\
        &= \frac{1}{n} \sum_{i=1}^n\sum_{s=1}^k \frac{1}{r} \sum_{j\in[n],s'\in[k]:\{v_i(s),v_j(s')\}\in\cE_t} \theta_{v_j(s')}\\
        &= \frac{1}{n}\sum_{j=1}^n \sum_{s'=1}^k \frac{1}{r} \sum_{i\in[n],s\in[k]:\{v_i(s),v_j(s)\}\in\cE_t} \theta_{v_j(s')}\\
        &= \frac{1}{n}\sum_{j=1}^n \sum_{s'=1}^k \theta_{v_j(s')} \frac{1}{r} \sum_{i\in[n],s\in[k]:\{v_i(s),v_j(s)\}\in\cE_t} \\
        &=  \frac{1}{n}\sum_{j=1}^n \sum_{s'=1}^k \theta_{v_j(s')} = \frac{1}{n}\sum_{j=1}^n  \theta_{j} = \bar\theta(\ell)\nonumber\\
    \end{align*}
    since  $\forall\,j\in[n],s'\in[k]$ we have $\sum_{i\in[n],s\in[k]:\{v_i(s),v_j(s)\}\in\cE_t}={\rm degree}(v_j(s'))=r$.
    
    For the second part, we can assume without loss of generality that $\bar\theta=0$, since this quantity is preserved by $\cC_t$. In that case, 
    \begin{align*}
        \mathbb{E}\left[\frac{1}{n}\sum_{i=1}^n\left\| \cC_t(\theta)_i - \bar\theta\right\|^2 \right]&=\frac{1}{n}\sum_{i=1}^n\sum_{\ell=1}^d\mathbb{E}\left[\left( \cC_t(\theta)_i(\ell)\right)^2 \right]\,.
    \end{align*}
    Let $\ell\in[d]$ be fixed.
        For all $i\in[n]$, let $v_i(s_\ell)$ be the chunk that includes coordinate $\ell$ of model chunks, for some $s_\ell\in[k]$ (without loss of generality, we can assume that this $s_\ell$ is the same for all $i$).
    We have:
    \begin{align*}
        \esp{\cC_t(\theta)_i(\ell)^2}&=\esp{ \left(\frac{1}{r} \sum_{s\in[k],j\in[n]:\{v_i(s),v_j(s_\ell)\}\in\cE_t}^n \theta_j(\ell)\right)^2 }\\
        &= \frac{1}{r^2}\sum_{j,j'\in[n]}\sum_{s,s'\in[k]} \theta_j(\ell)\theta_{j'}(\ell) \\
        &\times \proba{ \{v_j(s_\ell),v_i(s)\},\{v_{j'}(s_\ell),v_i(s')\}\in\cE_t}\,.
    \end{align*}
    For $j=j'$ and $s=s'$,
    \begin{align*}
        &\proba{ \{v_j(s_\ell),v_i(s)\},\{v_{j'}(s_\ell),v_i(s')\}\in\cE_t}\\
        &\qquad=\proba{ \{v_j(s_\ell),v_i(s)\}\in\cE_t}\\
        & \qquad= \frac{r}{nk-1}\,,
    \end{align*}
    which is the probability for a given edge of the virtual nodes graph to be in $\cE_t$, a graph sampled uniformly at random from all $r-$regular graphs.
    Let $j,j'\in[n]\setminus \{i\}$ and $s,s'\in[k]\setminus\{s_\ell\}$.
    If $j=j'$ but $s\ne s'$,
    \begin{align*}
        &\qquad\proba{ \{v_j(s_\ell),v_i(s)\},\{v_{j'}(s_\ell),v_i(s')\}\in\cE_t}\\
        &= \proba{ \{v_j(s_\ell),v_i(s)\}\in\cE_t\big|\{v_{j}(s_\ell),v_i(s')\}\in\cE_t}\times \frac{r}{nk-1}\,,
    \end{align*}
    and we thus need to compute $\proba{ \{v,w\}\in\cE_t\big|\{v,u\}\in\cE_t}$, for virtual nodes $u,v,w$ such that $u\ne w$.
    For all $w,w'\ne u$, by symmetry, we have $\proba{ \{v,w\}\in\cE_t\big|\{v,u\}\in\cE_t}=\proba{ \{v,w'\}\in\cE_t\big|\{v,u\}\in\cE_t}$.
    Then,
    \begin{align*}
        r&=\esp{{\rm degree} (v) \big| \{v,u\}\in\cE_t }\\
        & 1 + \sum_{w'\ne u,v} \proba{ \{v,w'\}\in\cE_t\big|\{v,u\}\in\cE_t}\\
        & 1 + (nk-2)\proba{ \{v,w\}\in\cE_t\big|\{v,u\}\in\cE_t}\,.
    \end{align*}
    Thus,
    \begin{align*}
        \proba{ \{v,w\}\in\cE_t\big|\{v,u\}\in\cE_t}= \frac{r-1}{nk-2}\,.
    \end{align*}
    By symmetry, we obtain the same result for $j\ne j'$, $s\ne s'$.
    Finally, we need to handle the case $j\ne j'$ and $s\ne s'$. This amounts to computing $\proba{ \{v,w\}\in\cE_t\big|\{v',w'\}\in\cE_t}$ for four virtual nodes $v,v',w,w'$ that are all different.
    We have:
    \begin{align*}
        r&=\esp{{\rm degree}(v) \big| \{v',w'\}\in\cE_t }\\
        & = (nk-3) \proba{ \{v,w\}\in\cE_t\big|\{v',w'\}\in\cE_t} \\
        &\quad + 2 \proba{ \{v,v'\}\in\cE_t\big|\{v',w'\}\in\cE_t}\,.
    \end{align*}
    From what we have already done, $\proba{ \{v,v'\}\in\cE_t\big|\{v',w'\}\in\cE_t}= \frac{r-1}{nk-2}$, leading to:
    \begin{align*}
        \proba{ \{v,w\}\in\cE_t\big|\{v',w'\}\in\cE_t}=\frac{r-2-\frac{r-1}{nk-2}}{nk-3}\,.
    \end{align*}
    Thus,
    \begin{align*}
        \esp{\cC_t(\theta)_i(\ell)^2}&= \frac{1}{r(nk-1)} \sum_{j\in[n],s\in[k]}\theta_j(\ell)^2\\
        &+\frac{1}{r(nk-1)} \sum_{j\in[n],s\ne s'\in[k]}\theta_j(\ell)^2 \frac{r-1}{nk-2}\\
        & + \frac{1}{r(nk-1)} \sum_{j\ne j'\in[n],s\in[k]}\theta_j(\ell)\theta_{j'}(\ell) \frac{r-1}{nk-2}\\
        & + \frac{1}{r(nk-1)} \sum_{j\ne j'\in[n],s\ne s'\in[k]} \theta_j(\ell)\theta_{j'}(\ell) \frac{r-2-\frac{r-1}{nk-2}}{nk-3}\\
        &= \frac{k}{r(nk-1)}\left(1 + \frac{(k-1)(r-1)}{(nk-2)}\right) \sum_{j\in[n]}\theta_j(\ell)^2 \\
        & + \frac{k}{r(nk-1)}\frac{r-1}{nk-2} \sum_{j\ne j'\in[n]}\theta_j(\ell)\theta_{j'}(\ell) \\
        & + \frac{k(k-1)}{r(nk-1)}  \frac{r-2-\frac{r-1}{nk-2}}{nk-3}\sum_{j\ne j'\in[n]} \theta_j(\ell)\theta_{j'}(\ell)\,.
    \end{align*}
    Then,
    \begin{align*}
        \sum_{j\ne j'\in[n]} \theta_j(\ell)\theta_{j'}(\ell)&=\left(\sum_{j\in[n]} \theta_j(\ell)\right)^2-\sum_{j\in[n]} \theta_j(\ell)^2\\
        &=-\sum_{j\in[n]} \theta_{j}(\ell)^2\,,
    \end{align*}
    leading to:
    \begin{align*}
        \esp{\cC_t(\theta)_i(\ell)^2}&= \frac{\rho}{n} \sum_{j\in[n]} \theta_{j}(\ell)^2\,,
    \end{align*}
    where 
    \begin{align*}
        \rho =\frac{nk}{r(nk-1)}\left(1+ \frac{(k-2)(r-1)}{nk-2}- \frac{(k-1)(r-2-\frac{r-1}{nk-2})}{nk-3}\right)\,,
    \end{align*}
    that satisfies
    \begin{align*}
        \rho \leq \frac{nk}{r(nk-1)}\left( 1+ \frac{k-r}{nk-3} + \frac{(k-1)(r-1)}{(nk-2)(nk-3)}\right)\,.
    \end{align*}
\end{proof}

Importantly, the inequality in \Cref{ineq:contraction} for $\rho$ as in \Cref{eq:rho} is \textit{almost tight}, up to lower order terms in the expression of $\rho$. 

Using this and adapting existing proofs for decentralized SGD with changing topologies and local updates \cite{pmlr-v119-koloskova20a,even2023unified} we enable ourself to prove the main result of \Cref{thm:convergence} to derive the formal convergence guarantees for \sys.

\begin{proof}[Proof of \Cref{thm:convergence}]
    The proof is based on the proofs of \cite[Theorem 2]{pmlr-v119-koloskova20a} and \cite[Theorem 4.2]{even2023unified}: thanks to their unified analyses, we only need to prove that their assumptions are verified for our algorithm.
    Our regularity assumptions are the same as theirs, we only need to satisfy the ergodic mixing assumption (\cite[Assumption 4]{pmlr-v119-koloskova20a} or \cite[Assumption 2]{even2023unified}).
    For $W_k\equiv\cC_{k/H}$ if $k\equiv 0 [H]$ and $W_k=I_n$ otherwise, we recover their formalism, and \cite[Assumption 4]{pmlr-v119-koloskova20a} is verified for $\tau=H$ and $p=1-\rho$.
    However, the proofs of \cite[Theorem 2]{pmlr-v119-koloskova20a} and \cite[Theorem 4.2]{even2023unified} require $W_k$ to be gossip matrices: this is not the case for our operators $\cC_t$. Still, $\cC_t$ can in fact be seen as gossip matrices on the set of \textit{virtual nodes}, and the only two properties required to prove \cite[Theorem 2]{pmlr-v119-koloskova20a} and \cite[Theorem 4.2]{even2023unified} are that $\cC_t$ conserves the mean, and the contraction proved in \Cref{lem:contraction}.
\end{proof}

\subsection{Proof of \Cref{th:prob_full_model}}\label{sec:proof:full_model_sharing}

\begin{proof}

For notational convenience, let us define the following: 

\begin{definition}[Distance between models]\label{def:dist_models}
    Let  $\sigma:\mathbb{R}\mapsto \{0,1\}$ denote the \emph{discrete metric}, \ie, $\forall\,z,z'\in\mathbb{R}$,
    $\sigma(z,z')=\begin{cases}
       0 \text{, if $z=z'$,}\\
       1 \text{, otherwise}
   \end{cases}.$
   Then, for a pair of models $\theta,\theta'\in\mathbb{R}^d$, let the \emph{distance} between them be $\delta(\theta,\theta')$ such that $\delta(\theta,\theta')=\sum\limits_{p=1}^d\sigma(\theta(p),\theta'(p))$.
\end{definition}

We note that $\delta\left(\theta^{(t)}_j,\theta^{(t)}_{i\leftarrow j}\right)=\left\lvert \theta^{(t)}_{i\not\leftarrow j}\right\rvert$. Let us fix \acp{RN} $N_i$ and $N_j$ where $i,j\in[n]$ and $i\neq j$. Assuming that each \ac{VN} has a degree of $r$, when there are $k$ \acp{VN} per \ac{RN}, $\delta\left(\theta^{(t)}_j,\theta^{(t)}_{i\leftarrow j}\right)=0$ can occur only when all $k$ \acp{VN} of $N_j$ independently communicate with at least one of $k$ \acp{VN} of $N_i$ in the given communication round. Therefore, we have:
    \begin{align}
        &\mathbb{P}\left[\delta\left(\theta^{(t)}_j,\theta^{(t)}_{i\leftarrow j}\right)=0\right]=\left(1-\left(1-\frac{r}{nk-1}\right)^k\right)^k\nonumber
    \end{align}
    In order to prove the desired result, we need to show:
    \begin{align}
        &\left(1-\left(1-\frac{r}{n(k+1)-1}\right)^{k+1}\right)^{k+1}\nonumber\\
        &\leq\left(1-\left(1-\frac{r}{nk-1}\right)^k\right)^k\label{eq:prob_full_model_goal1}
    \end{align}

In order to show \eqref{eq:prob_full_model_goal1}, we will first show 
\begin{align}
     &\left(1-\frac{r}{nk-1}\right)^k\leq\left(1-\frac{r}{n(k+1)-1}\right)^{k+1}\label{eq:func_intK}
\end{align}

To this, we define an auxiliary function $F\colon[1,\infty)\mapsto \mathbb{R}$ such that $F(x)=\left(1-\frac{r}{nx-1}\right)^x$. Hence, for \eqref{eq:prob_full_model_goal1} to hold, it suffices to show that $F(.)$ is a non-decreasing function of $x$. In other words, for every $x\in[1,\infty)$, we wish to show:
    \begin{align}
        &\frac{d}{dx}F(x)\geq 0\nonumber\\
        \iff&\left(\frac{nx-1-r}{{n x-1}}\right)^x \left( \frac{n r x}{{(n x - 1)(n x - r - 1)}}\right.\nonumber\\ 
        &\left.+ \ln\left(\frac{nx-1-r}{{n x-1}} \right) \right)\geq 0\nonumber
    \end{align}

Noting that $\left(1- \frac{r}{{n x-1}}\right)^x>0$, it suffices to show:
    \begin{align}
        &\frac{n r x}{{(n x - 1)(n x - r - 1)}} + \ln\left(1- \frac{r}{{n x}-1}\right) \geq 0\nonumber\\
        \iff&\frac{n r x}{{(n x - 1)(n x - r - 1)}}-\ln\left(\frac{nx-1}{nx-r-1}\right)\geq 0\nonumber\\
        \iff&\frac{n r x}{{(n x - 1)(n x - r - 1)}}\geq\ln\left(\frac{nx-1}{nx-r-1}\right)\label{eq:aux_deriv_1}
    \end{align}
Recalling that $i)$ $\ln(z)\leq z-1$ $\forall\, z\geq 0$, $ii)$ $n>1$, $n>1$, and $iii)$ $r< nx-1$, we get: 
\begin{align}
    &\ln\left(\frac{nx-1}{nx-r-1}\right)\leq \frac{nx-1}{nx-r-1}-1 = \frac{r}{nx-r-1}\label{eq:aux_deriv_2}  
\end{align}
Thus, incorporating \eqref{eq:aux_deriv_2} into \eqref{eq:aux_deriv_1}, it suffices to show that:
\begin{align}
    &\frac{n r x}{{(n x - 1)(n x - r - 1)}}\geq\frac{r}{nx-r-1}\nonumber\\
    \iff&\frac{nx}{nx-1}\geq 1, \text{ which trivially holds.}\nonumber
\end{align}
    Hence, we establish that:
    \begin{align}
        &\left(1-\frac{r}{nk-1}\right)^k\leq\left(1-\frac{r}{n(k+1)-1}\right)^{k+1}\nonumber\\
        \implies &1-\left(1-\frac{r}{n(k+1)-1}\right)^{k+1} \nonumber\\
        &\leq 1-\left(1-\frac{r}{nk-1}\right)^k\nonumber\\
        \implies& \left(1-\left(1-\frac{r}{n(k+1)-1}\right)^{k+1}\right)^k \nonumber\\
        &\leq \left(1-\left(1-\frac{r}{nk-1}\right)^k\right)^k\label{eq:prob_full_model_goal2}
    \end{align}
    But $1-\left(1-\frac{r}{n(k+1)-1}\right)^{k+1}\leq 1$ implies:
    \begin{align}
        &\left(1-\left(1-\frac{r}{n(k+1)-1}\right)^{k+1}\right)^{k+1}\nonumber\\
        &\leq \left(1-\left(1-\frac{r}{n(k+1)-1}\right)^{k+1}\right)^k\label{eq:prob_full_model_goal3}
    \end{align}
    Therefore, combining \eqref{eq:prob_full_model_goal2} and \eqref{eq:prob_full_model_goal3}, we obtain \eqref{eq:prob_full_model_goal1}, as desired. 
\end{proof}

\subsection{Proof of \Cref{th:expected_params}}\label{sec:proof:expected_params}

\begin{proof}
    Setting $S$ as the random variable denoting the number of \acp{VN} of $N_j$ that connect with at least one of the \acp{VN} of $N_i$ (i.e., the \acp{VN} of $N_j$ that are responsible for sharing the corresponding chunks of $N_j$'s model they hold with $N_i$) and, hence, noting that the number of parameters of $N_j$'s model that are shared with $N_i$ is $\frac{d}{k}S$, we essentially need to show that $\mathbb{E}\left[\delta\left(\theta^{(t)}_j,\theta^{(t)}_{i\leftarrow j}\right)\right]=\mathbb{E}\left[\frac{d}{k}S\right]$ is a decreasing function of $k$, where $\delta(.)$ is same as has been defined in \Cref{def:dist_models}. By the \emph{law of the unconscious statistician}, we have:
    \begin{align}
        &\mathbb{E}\left[\frac{dS}{k}\right]=\sum_{s=0}^k\frac{ds}{k}\mathbb{P}\left[S=s\right]\nonumber\\
        &=\sum_{s=0}^k\frac{ds}{k}{k\choose s}\left(1-(1-p)^k\right)^{s}(1-p)^{k(k-s)}\nonumber\\
        &\left[\text{where } p=\frac{r}{nk-1}\right]\nonumber\\
        &=\frac{d}{k}\sum_{s=0}^ks{k\choose s}\left(1-(1-p)^k\right)^{s}(1-p)^{k(k-s)}\nonumber\\
        &=\frac{d}{k}\mathbb{E}_{X\sim \operatorname{Bin}(k,\pi_k)}(X)\nonumber\\
        &\left[\text{where } \pi_k=1-\left(1-\frac{r}{nk-1}\right)^k\right]\nonumber\\
        &=\frac{d}{k}k\pi_k=d\left(1-\left(1-\frac{r}{nk-1}\right)^k\right)
    \end{align}

    By \eqref{eq:prob_full_model_goal2} in \Cref{th:prob_full_model}, we know that $\pi_k=1-\left(1-\frac{r}{nk-1}\right)^k$ is a decreasing function of $k$. Hence, for a fixed $n$, $d$, and $r$, $\mathbb{E}\left[\frac{dS}{k}\right]=d\pi_k$ is a decreasing function of $k$. 
\end{proof}

\subsection{Proof of \Cref{th:mutual_info}}\label{sec:proof:mutual_info}

\begin{definition}[Mutual information\cite{ShannonInfoTheory1948}]
\label{def:MI}
Let $(X,Y)$ be a pair of random variables defined over the discrete space $\mathcal{X}\times\mathcal{Y}$ such that $p_{XY}$ is the joint PMF of $X$ and $Y$, $p_{X}$ and $p_{Y}$ are the corresponding marginal PMFs, and $p_{X|Y}$ is the conditional probability of $X$ given $Y$. Then the (Shannon) \emph{entropy} of $X$, $H(X)$, is defined as $H(X)= -\sum\limits_{x\in\mathcal{X}}p_X(x) \log p_X(x)$. The \emph{residual  entropy} of $X$ given $Y$ is defined as $H(X|Y)= \sum\limits_{y\in\mathcal{Y}}p_Y(y) H(X|Y=y) =  -\sum\limits_{y\in\mathcal{Y}}p_Y(y)\sum\limits_{x\in\mathcal{X}}p_{X|Y}p(x|y) \log p_{X|Y}(x|y)$, and, finally, the 
\emph{\ac{MI}} is given by: $$I(X;Y)=H(X) - H(X|Y) =\sum\limits_{x\in\mathcal{X}}\sum\limits_{y\in\mathcal{Y}}p_{XY}(x,y)\ln\frac{p_{XY}(x,y)}{p_{X}(x)p_{Y}(y)}$$
\end{definition}

\begin{proof}
\Cref{assump:MI_1} implies that $X\sim\mathcal{N}\left(\vb*{\mu}^{(t)}_{X_{ij}},\Sigma^{(t)}_{X_{ij}}\right)$ and $Y\sim\mathcal{N}\left(\vb*{\mu}^{(t)}_{Y_{ij}},\Sigma^{(t)}_{Y_{ij}}\right)$, where $\vb*{\mu}^{(t)}=\left(\vb*{\mu}^{(t)}_{X_{ij}},\vb*{\mu}^{(t)}_{Y_{ij}}\right)^{\operatorname{T}}$. %
Hence, 
    \begin{align}
    &I\left(X_{ij};Y_{ij}\right)=D_{\operatorname{KL}}\left(p(X,Y)\left\lvert\right\rvert p(X)p(Y)\right)\nonumber\\
    &=D_{\operatorname{KL}}\left(\mathcal{N}\left(\vb*{\mu}^{(t)}_j,\Sigma^{(t)}_j\right)\left\lvert\right\rvert \mathcal{N}\left(\hat{\vb*{\mu}}^{(t)}_j,\hat{\Sigma}^{(t)}_j\right)\right)\nonumber\\
    &\left[\text{where } \hat{\vb*{\mu}}^{(t)}_j=\vb*{\mu}^{(t)}_j,\,\hat{\Sigma}^{(t)}_j=\begin{pmatrix}
\Sigma^{(t)}_{X_{ij}} & \vb*{0}\nonumber\\
\vb*{0} & \Sigma^{(t)}_{Y_{ij}}
\end{pmatrix}\right]\\
    &=\frac{1}{2}\ln\left(\frac{\operatorname{det}\left(\Sigma^{(t)}_{X_{ij}}\right)\operatorname{det}\left(\Sigma^{(t)}_{Y_{ij}}\right)}{\operatorname{det}\left(\Sigma^{(t)}_j\right)}\right)\nonumber\\
    &=\frac{1}{2}\ln\left(\frac{\operatorname{det}\left(\Sigma^{(t)}_{X_{ij}}\right)}{\operatorname{det}\left(\Sigma^{(t)}_{X_{ij}}-\Sigma^{(t)}_{X_{ij}Y_{ij}}\Sigma^{(t)-1}_{Y_{ij}}\Sigma^{(t)}_{Y_{ij}X_{ij}}\right)}\right)\label{eq:MI_step1}\\
    &\left[\because\,\operatorname{det}\left(\Sigma^{(t)}_j\right)\right.\nonumber\\
    &\left.=\operatorname{det}\left(\Sigma^{(t)}_{Y_{ij}}\right)\operatorname{det}\left(\Sigma^{(t)}_{X_{ij}}-\Sigma^{(t)}_{X_{ij}Y_{ij}}\Sigma^{(t)-1}_{Y_{ij}}\Sigma^{(t)}_{Y_{ij}X_{ij}}\right)\right]\nonumber
    \end{align}
    Recalling \Cref{assump:MI_2}, by Hadamard's theorem on determinants~\cite{Hadamard1893determinants,Hadamard1893determinantsWiki}, we note that 
    \begin{align}
        &\operatorname{det}\left(\Sigma^{(t)}_{X_{ij}}\right)\leq B^{d(t)}d(t)^{d(t)/2}\label{eq:MI_HadamardIneq}
    \end{align}
    Moreover, observing that $\Sigma^{(t)}_{X_{ij}}-\Sigma^{(t)}_{X_{ij}Y_{ij}}\Sigma^{(t)-1}_{Y_{ij}}\Sigma^{(t)}_{Y_{ij}X_{ij}}$ %
    is the Schur complement $\Sigma^{(t)}_j/\Sigma^{(t)}_{Y_{ij}}$ of the block $\Sigma^{(t)}_{Y_{ij}}$ in $\Sigma^{(t)}_j$,  under Assumptions \ref{assump:MI_3} and \ref{assump:MI_4} we use the determinantal lower bounds derived by Kalantari and Pate (Corollary 2 of \cite{kalantarideterminantsbound2001}) to obtain
    \begin{align}
        &\operatorname{det}\left(\Sigma^{(t)}_j/\Sigma^{(t)}_{Y_{ij}}\right)\geq \alpha^{\kappa}\beta^{d(t)-\kappa}\label{eq:MI_KalantariIneq}
    \end{align}
    where
    \begin{align}
        &\kappa=\frac{\beta d(t)-\operatorname{tr}\left(\Sigma^{(t)}_j/\Sigma^{(t)}_{Y_{ij}}\right)}{\beta-\alpha}.\nonumber
    \end{align}

    Incorporating \eqref{eq:MI_HadamardIneq} and \eqref{eq:MI_KalantariIneq} into \eqref{eq:MI_step1}, and setting $\Gamma$ and $\hat{B}$ as defined in the statement of the theorem, we conclude $I(X;Y)\leq \Gamma \hat{B}^{d(t)} d(t)^{d(t)/2}$, as required.

\end{proof}

\end{document}